\documentclass[aps,prl,showpacs,superscriptaddress,twocolumn]{revtex4-1}

\usepackage{amsfonts}
\usepackage{subfigure}
\usepackage{amsmath}
\usepackage{txfonts}
\usepackage{amssymb}
\usepackage{amsbsy} %for sigma in bold face
\usepackage{epsfig}
\usepackage{graphicx}
\usepackage{epstopdf}

\usepackage{mathdots} %from Fei Song

\def\be{\begin{equation}} \def\ee{\end{equation}}
\def\bea{\begin{eqnarray}} \def\eea{\end{eqnarray}}

\def\nn{\nonumber}

\def\bk{{\bf k}}

\def\bx{{\bf x}}

\def\be{{\bf e}}

\newcommand{\bra}[1]{\langle#1|}
\newcommand{\ket}[1]{|#1\rangle}

\def\rw{\rightarrow}

\def\la{\langle}
\def\ra{\rangle}

\begin{document}

\title{Non-Hermitian Chern bands}

%\title{Bulk-boundary correspondence of non-Hermitian Chern insulators}

\author{Shunyu Yao}
 \affiliation{ Institute for
Advanced Study, Tsinghua University, Beijing,  100084, China }

\author{Fei Song}
 \affiliation{ Institute for
Advanced Study, Tsinghua University, Beijing,  100084, China }

\author{Zhong Wang} \altaffiliation{ wangzhongemail@gmail.com }
\affiliation{ Institute for
Advanced Study, Tsinghua University, Beijing,  100084, China }

\affiliation{Collaborative Innovation Center of Quantum Matter, Beijing, 100871, China }

%\date{\today}

\begin{abstract}

The relation between chiral edge modes and bulk Chern numbers of quantum Hall insulators is a paradigmatic example of bulk-boundary correspondence. We show that the chiral edge modes are not strictly tied to the Chern numbers defined by a non-Hermitian Bloch Hamiltonian. This breakdown of conventional bulk-boundary correspondence stems from the non-Bloch-wave behavior of eigenstates (non-Hermitian skin effect), which generates pronounced deviations of phase diagrams from the Bloch theory. We introduce non-Bloch Chern numbers that faithfully predict the numbers of chiral edge modes. The theory is backed up by the open-boundary energy spectra, dynamics, and phase diagram of representative lattice models. Our results highlight a unique feature of non-Hermitian bands and suggest a non-Bloch framework to characterize their topology.

\end{abstract}

\maketitle

Hamiltonians are Hermitian in the standard quantum mechanics. Nevertheless, non-Hermitian Hamiltonians\cite{bender2007making,bender1998real} are highly useful in describing many phenomena such as various open systems\cite{rotter2009non,malzard2015open, carmichael1993,zhen2015spawning, diehl2011topology,cao2015microcavities,choi2010coalescence, san2016majorana,lee2014heralded,lee2014entanglement} and waves propagations with gain and loss\cite{makris2008beam,longhi2009bloch,ruter2010observation,klaiman2008branch,bittner2012,regensburger2012parity,
guo2009complex,liertzer2012pumpinduced,peng2014lossinduced,lin2011unidirectional,Lu2014review,
feng2013experimental,fleury2015invisible,chang2014PT,hodaei2017enhanced, hodaei2014PT,feng2014singlemode,kawabata2017retrieval,gao2015billiard,xu2016topological, ashida2017parity,chen2017exceptional,ding2016multiple,downing2017, ozawa2018rmp,el2018non,longhi2018}. Recently, topological phenomena in non-Hermitian systems have attracted considerable attention. For example, an electron's non-Hermitian self energy stemming from disorder scatterings or electron-electron interactions\cite{kozii2017,papa2018bulk,shen2018quantum} can generate novel topological effects such as bulk Fermi arcs connecting exceptional points\cite{kozii2017,papa2018bulk} (a photonic counterpart has been observed experimentally\cite{Zhou2018arc}). The interplay between non-Hermiticity and topology has been a growing field with a host of interesting theoretical\cite{shen2017topological, esaki2011,lee2016anomalous,leykam2017,lieu2018ssh,yin2018ssh,menke2017,rudner2009topological,li2017kitaev,xiong2017,
alvarez2017,gong2018nonhermitian,yao2018edge,liang2013topological, hu2011absence,gong2010geometrical,rudner2016survival,zhu2014PT,gong2017zeno,wang2015spontaneous,
kawabata2018PT,ni2018exceptional,zyuzin2018flat,cerjan2018weyl, klett2017sshkitaev,zhou2017dynamical,gonzalez2017,yuce2016majorana,hu2017exceptional, xu2017weyl,ke2017topological,harari2018topological} and experimental\cite{zeuner2015bulk,xiao2017observation, weimann2017topologically,poli2015selective,parto2017SSHexperiment,zhao2017Topological,zhan2017detecting} progresses witnessed in recent years.

A central principle of topological states is
the bulk-boundary (or bulk-edge) correspondence, which asserts that the robust boundary states are tied to the bulk topological invariants. Within the band theory, the bulk topological invariants are defined using the Bloch Hamiltonian\cite{hasan2010,qi2011,Chiu2016rmp, bernevig2013topological}. This has been well understood in the usual context of Hermitian Hamiltonians; nevertheless, it is a subtle issue to generalize this correspondence to non-Hermitian systems\cite{shen2017topological,esaki2011,lee2016anomalous,leykam2017,xiong2017, lieu2018ssh,alvarez2017,gong2018nonhermitian,yao2018edge}. As demonstrated numerically\cite{lee2016anomalous,xiong2017,alvarez2017,yao2018edge}, the bulk spectra of one-dimensional (1D) open-boundary systems dramatically differ from those with periodic boundary condition, suggesting a breakdown of bulk-boundary correspondence. This issue has been resolved\cite{yao2018edge} in 1D non-Hermitian Su-Schrieffer-Heeger (SSH) model: The topological end modes are determined by the non-Bloch winding number\cite{yao2018edge} instead of topological invariants defined by Bloch Hamiltonian\cite{esaki2011,lee2016anomalous,leykam2017, rudner2009topological,lieu2018ssh, menke2017,yin2018ssh,li2017kitaev}, which suggests a generalized bulk-boundary correspondence\cite{yao2018edge}.

However, the general implications of these results based solely on a simple 1D model remain to be understood (e.g., Is the physics specific to 1D?). Moreover, the topology of this 1D model requires a chiral symmetry\cite{Chiu2016rmp}, which is often fragile in real systems. Thus, we are motivated to study 2D non-Hermitian Chern insulators whose robustness is independent of symmetries\cite{thouless1982,Haldane1988,chang2013experimental,wang2009observation}. In addition, non-Hermitian Chern bands are relevant to a number of physical systems (e.g. photonic Chern insulators\cite{ozawa2018rmp} with gain/loss, topological-insulator lasers\cite{harari2018topological,bandres2018topological}, interacting/disordered electron systems\cite{kozii2017}). They have been characterized by non-Hermitian generalizations of Bloch Chern numbers\cite{shen2017topological,esaki2011}, which are expected to predict the edge states.

In this paper, we uncover an unexpected bulk-boundary correspondence of non-Hermitian Chern bands. We find that the chiral edge states are not strictly related to the Chern numbers of non-Hermitian Bloch Hamiltonians. More remarkably, in spite of this breakdown of conventional bulk-boundary correspondence, the edge states retain a general topological characterization. In fact, the ``breakdown'' stems from the general non-Bloch-wave behavior of eigenstates (non-Hermitian skin effect), which affects the phase diagrams in a dramatic yet predictable manner. We therefore introduce ``non-Bloch Chern numbers'' to which the numbers of chiral edge modes are strictly tied. Notably, complex-valued wavevector (momentum) is used in their construction, which captures a unique feature of non-Hermitian bands. As an illustration, we study a concrete lattice model, whose energy spectra, dynamics (edge wave propagations), and phase diagram is found to be in accordance with our theory.

\emph{Bloch Hamiltonian.--}We consider a lattice model similar to that of Ref.\cite{shen2017topological}. The Bloch Hamiltonian is \bea H(\bk)&=& (v_x\sin k_x+i\gamma_x)\sigma_x+  (v_y\sin k_y+i\gamma_y)\sigma_y \nn\\ && +(m  -t_x\cos k_x -t_y\cos k_y+i\gamma_z)\sigma_z, \label{model} \eea  where $\sigma_{x,y,z}$ are Pauli matrices. The Hermitian part is the Qi-Wu-Zhang model\cite{qi2005}(a variation of Haldane model\cite{Haldane1988}); the non-Hermitian parameters $\gamma_{x,y,z}$ appear as ``imaginary Zeeman fields''\cite{lee1952}. When $\gamma_{x,y,z}=0$, the model has a topological transition at $m =t_x+t_y$, where the Chern number jumps. We shall focus on $m$ being close to $t_x+t_y$ ($\gamma_{x,y,z}$ are taken to be small compared to $t_{x,y}$). The eigenvalues of $H(\bk)$ are \bea E_\pm(\bk)=\pm \sqrt{\sum_{j=x,y,z} (h_j^2 -\gamma_j^2  +2i \gamma_j h_j )}, \label{blochE} \eea where $(h_x,h_y,h_z)=(v_x\sin k_x, v_y\sin k_y, m -\sum_jt_j\cos k_j)$.

A band is called ``gapped'' or ``separable''\cite{shen2017topological} if its energies in the complex plane are separated from those of other bands. In this model, the Bloch bands are gapped if $E_\pm(\bk)\neq 0$. The gapped regions are found to be $m>m_+$ and $m <m_-$, where $m_\pm$ have simple expressions when $\gamma_z=0$:
\bea m_\pm =t_x + t_y  \pm\sqrt{\gamma_x^2+\gamma_y^2}. \label{mpm} \eea The Bloch phase boundaries are $m=m_\pm$, where the gap closes at $\bk=(0,0)$. One can obtain that the $H(\bk)$-based Chern number (Bloch Chern number)\cite{shen2017topological,esaki2011} is $0$ for $m >m_+$,  $1$ for $m <m_-$, and becomes non-definable in the gapless region $m \in [m_-,m_+]$.

\emph{Open boundary.--}According to the usual bulk-boundary-correspondence scenario, the chiral edge states of an open-boundary system should be determined by the Bloch Chern numbers. However, a different physical picture is found here. Let us present numerical results before the theory. To be concrete, let us take $\gamma_z=0$ and focus on the $x$-$y$-symmetric cases, namely $v_x=v_y=v, t_x=t_y=t, \gamma_x=\gamma_y=\gamma$. We fix $v=t=1$ and solve the real-space lattice Hamiltonian on a square geometry with edge length $L$ in both $x$ and $y$ directions, taking $(m,\gamma)$ as the varied parameters.  Among other results we find:

(i) The open-boundary spectra are prominently different from those of Bloch Hamiltonian. Although the Bloch spectra are complex-valued [see Eq.(\ref{blochE})], the majority of square-geometry energy eigenvalues are real-valued when $\gamma_z=0$.  It should be mentioned that the reality of open-boundary spectra is not a general rule; in other models, they are often complex (e.g., when $\gamma_z$ is nonzero); nevertheless, in general the open-boundary and Bloch spectra have pronounced differences.

The reality of square-geometry spectra can be explained as follows. To avoid lengthy notations, we simply take $L=2$ as an illustration. Let us order the four sites as $(x,y)=(1,1), (2,1), (1,2), (2,2)$, then the real-space Hamiltonian reads
\bea
H=
\begin{pmatrix}
M&T_x&T_y&0 \\T_{x}^{\dag}&M&0&T_y\\T_y^{\dag}&0&M&T_x \\0&T_y^{\dag} &T_x^{\dag} &M
\end{pmatrix}
\eea
where
\bea
M =&& m \sigma_z+i\gamma_x\sigma_x+i\gamma_y\sigma_y,\nonumber\\
T_x =-\frac{t_x}{2}\sigma_z- && i\frac{v_x}{2}\sigma_x, \quad
T_y=-\frac{t_y}{2}\sigma_z-i\frac{v_y}{2}\sigma_y.
\eea
This Hamiltonian is ``$\eta$-pseudo-Hermitian''\cite{Mostafazadeh2002i,Mostafazadeh2002ii} (not $PT$-symmetric\cite{el2018non,hu2011absence}), namely, it satisfies $\eta^{-1}H^{\dag}\eta=H$, where $\eta$ is the direct product of spatial inversion and $\sigma_z$:
\bea \eta=
\begin{pmatrix}
0&0&0 &\sigma_z\\0&0 &\sigma_z&0\\0 &\sigma_z&0&0 \\\sigma_z&0&0&0
\end{pmatrix}.
\eea     The pseudo-Hermiticity guarantees that from $H\ket{\psi_n}=E_n\ket{\psi_n}$, one can infer $E_n\bra{\psi_n}\eta\ket{\psi_n}=E^*_n\bra{\psi_n}\eta\ket{\psi_n}$, which means $E_n=E_n^*$ when $\bra{\psi_n}\eta\ket{\psi_n}\neq 0$. In this model, we find that the majority of eigenstates have $\bra{\psi_n}\eta\ket{\psi_n}\neq 0$.

The dissimilarity between open-boundary and Bloch spectra has also been found in a 1D model\cite{lee2016anomalous,xiong2017,alvarez2017,yao2018edge},  whose spectra can be readily obtained via a similarity transformation to a Hermitian Hamiltonian\cite{yao2018edge}. Free of this specificity, our 2D model is a more nontrivial and representative exemplification of the phenomenon that the open-boundary spectra are noticeably different from the Bloch spectra.

\begin{figure}
\includegraphics[width=8cm, height=5.0cm]{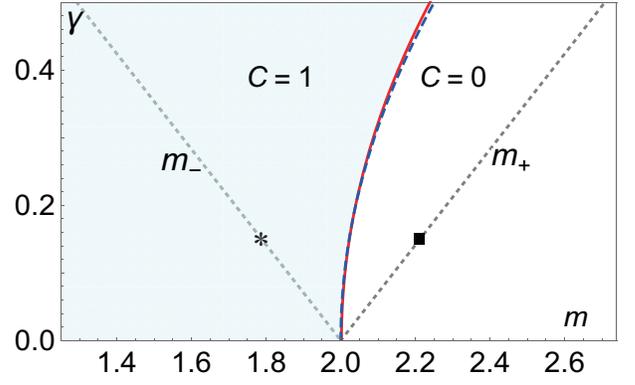}
\caption{ Topological phase diagram based on open-boundary spectra (for $v_{x,y}=t_{x,y}=1,\, \gamma_{x,y}=\gamma,\,\gamma_z=0$). Chiral edge states are found in the shadow area, which is therefore topologically nontrivial. The trivial-nontrivial phase boundary (red solid curve) is well approximated by the theoretical curve in Eq.(\ref{quadratic}) (shown as the blue dashed curve, which is very close to the red solid curve). Away from this phase boundary, the (open-boundary) bulk spectra are gapped. The Bloch-Hamiltonian phase boundaries are shown as the dotted lines, whose equations are $m =m_\pm$ with $m_\pm=2\pm\sqrt{2}\gamma$. The Bloch spectra are gapless in the fan $m \in[m_-,m_+]$. The non-Bloch Chern number $C$ is defined in Eq.(\ref{chern}) (We take the $\text{Re}(E_\alpha)<0$ band and omit the $\alpha$ index; see text). } \label{phase}
\end{figure}

(ii) The topological transition between nontrivial and trivial phases (i.e., with and without robust chiral edge modes) does not occur at the Bloch phase boundary $m=m_\pm$ [Eq.(\ref{mpm})]. By numerically scanning the gap-closing points\footnote{The gap in $L\rw\infty$ limit is determined from the intercept in the $\text{gap}^2$-$1/L^2$ plot.}, we find that the phase boundary is a single curve (red solid one in Fig.\ref{phase}), in sharp contrast to the two straight lines $m =m_\pm$ obtained from the Bloch Hamiltonian. Furthermore, the numerical phase boundary can be well approximated by the theoretical prediction of Eq.(\ref{quadratic}).

\begin{figure*}[htb]
\subfigure{\includegraphics[width=17cm, height=3.7cm]{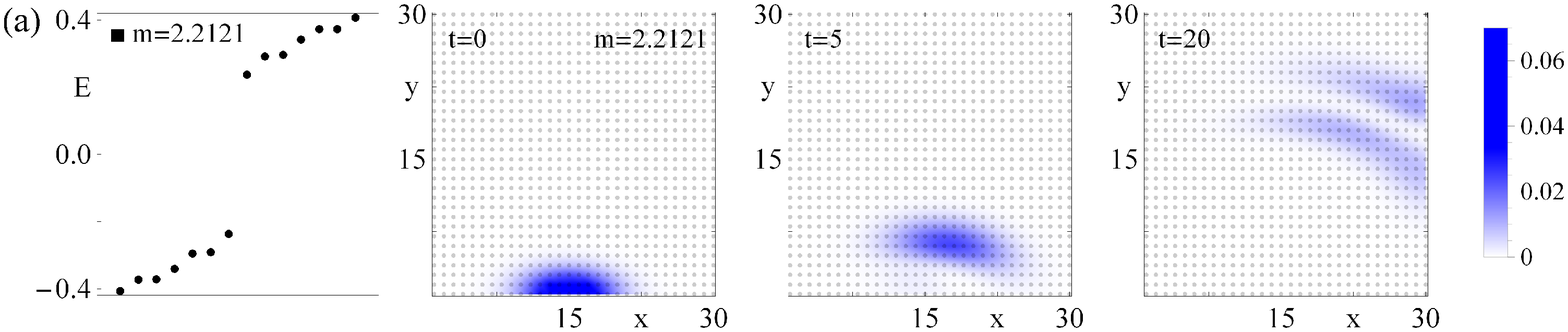}}
\subfigure{\includegraphics[width=17cm, height=3.7cm]{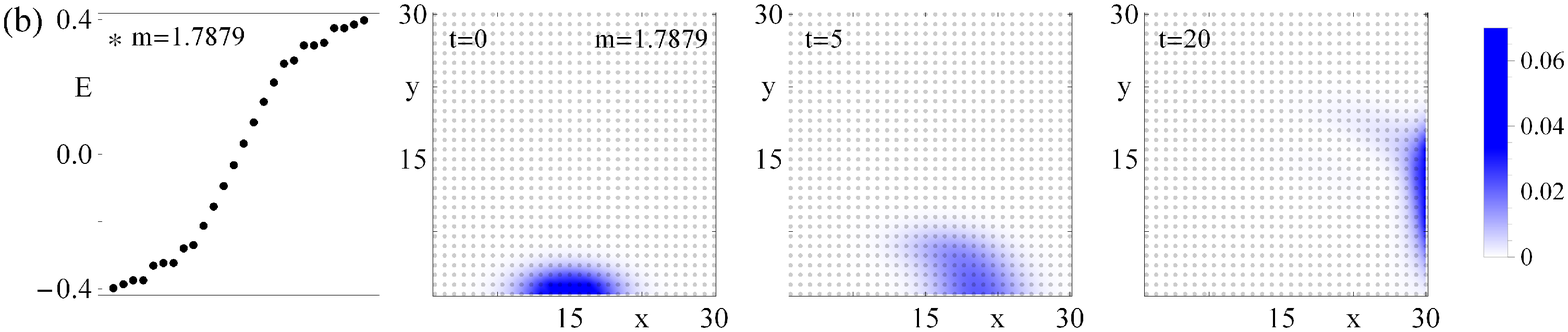}}
\caption{ Left panel: Lowest energy eigenvalues of a square geometry with $L=30$. Right three panels: wave packet evolutions. (a) $m=2.2121$; (b) $m=1.7879$ (indicated by $\blacksquare$ and $\ast$ in Fig.\ref{phase}), with $\gamma=0.15$ for both. The energy eigenvalues shown here are real-valued. In (a), a nonzero energy gap is apparent; in (b), there are a few in-gap energies of chiral edge states. For the wave-packet evolution, the initial state takes the Gaussian form $\psi(t=0)=\mathcal{N}\exp[-(x-15)^2/40-(y-1)^2/10] (1,1)^T$, $\mathcal{N}$ being the normalization factor, and evolves according to the Schrodinger equation $i\partial_t\ket{\psi(t)}=H\ket{\psi(t)}$. The intensity profile of $\ket{\psi(t)}$ (modulus squared), normalized so that the total intensity is $1$, is shown for $t=0,5,20$. The wave packet quickly fades into the bulk in (a), while the chiral (unidirectional) edge motion is appreciable in (b). }\label{square}
\end{figure*}

As an illustration of the phase diagram, we show in Fig. \ref{square} the numerical spectra for two values of parameters indicated as
$\blacksquare$ and $\ast$ in Fig.\ref{phase}.  Both $\blacksquare$ and $\ast$ are taken at the Bloch phase boundary where the Bloch Hamiltonian is gapless. Remarkably, the spectra at $\blacksquare$ clearly display an energy gap $\approx 0.4$. A similar bulk gap is found for the $\ast$ point; in addition, there are a few in-gap energies, which can be identified as those of chiral edge modes. The absence/existence of chiral edge modes can also be detected by wave packet motions (Fig.\ref{square}, right panels). In Fig.\ref{square}(a), there is no chiral edge mode, and the initial wavefunction are superpositions of bulk eigenstates, therefore the wave packet quickly enters the bulk; in Fig.\ref{square}(b), one can see clear signatures of chiral motions along the edge.

Finally, we emphasize that the phase diagram is independent of the geometry of system, which indicates its topological nature. For example, the disk geometry ($x^2+y^2\leq R^2$) produces the same phase diagram as Fig.\ref{phase}\cite{supplemental}.

\emph{Non-Bloch Chern number.--}This intriguing phase diagram is a prediction of the non-Bloch theory based on complex-valued wavevectors. We now introduce this formulation. First, we find that all the bulk eigenstates are exponentially localized at the boundary of system\footnote{It remains meaningful to talk about ``bulk eigenstates'' because their number grows as $L^2$}\footnote{In 1D non-Hermitian SSH models, a similar phenomenon has been seen numerically\cite{alvarez2017} and analytically\cite{yao2018edge}. Our 2D model establishes that this phenomenon is not merely a peculiar 1D effect but a general feature of non-Hermitian bands.}. This ``non-Hermitian skin effect''\cite{yao2018edge} is possible because the eigenstates are non-orthogonal. To see this effect explicitly, we consider the low-energy continuum model of Eq.(\ref{model}) (with $\gamma_z=0$), which is its expansion to the $k_j^2$ order:
\bea H(\bk)= (v_x k_x+i\gamma_x)\sigma_x+ (v_yk_y+i\gamma_y) \sigma_y  \nn \\ +(m-t_x-t_y +\frac{t_x}{2}k_x^2 +\frac{t_y}{2}k_y^2  )\sigma_z. \quad   \eea It can be decomposed as $H(\bk)=H_0+H_1$, where $H_1=i\gamma_x\sigma_x+i\gamma_y\sigma_y$, and $H_0$ is the rest part. For small $\bk$, we have $\frac{\partial H_0}{\partial k_j}=v_j\sigma_j$ and $H_1=i\sum_{j=x,y} \frac{\gamma_j}{v_j}\frac{\partial H_0}{\partial k_j}= \sum_j\frac{\gamma_j}{v_j}[x_j,H_0]$, where $(x_x,x_y)\equiv (x,y)$. Note that $x_j =i\frac{\partial}{\partial k_j}$ in the $\bk$-space representation. Let us treat $H_1$ as a perturbation. The lowest-order perturbation to an eigenstate $\ket{n}$ of $H_0$ is $\sum_{l\neq n}\frac{\ket{l}\bra{l}H_1\ket{n}}{E_n-E_l} =\sum_{l\neq n} \sum_j\frac{\gamma_j}{v_j}\ket{l}\bra{l}x_j\ket{n} =\sum_j\frac{\gamma_j}{v_j}(x_j-\bar{x}_j)\ket{n}$, where $\bar{x}_j\equiv \bra{n}x_j\ket{n}$. Therefore, the associated eigenstate of $H$ is $\ket{\psi_n}=[1+\sum_j\frac{\gamma_j}{v_j}(x_j-\bar{x}_j)]\ket{n} \approx \exp[\sum_j\frac{\gamma_j}{v_j}(x_j-\bar{x}_j)]\ket{n}$. Thus, for an extended state $\ket{n}$, $\ket{\psi_n}$ is exponentially localized like $\exp[(\gamma_x/v_x) x+(\gamma_y/v_y) y]$\footnote{Exponential localization is also confirmed numerically\cite{supplemental}.}. The role of non-Hermiticity is notable in this derivation: Without the ``$i$'' factor in $H_1$, we would have obtained a phase factor instead of exponential decay.

Because of this non-Hermitian skin effect, we take a complex-valued wavevector (or momentum) to describe open-boundary eigenstates: \bea  \bk \rw \tilde{\bk} +i \tilde{\bk}', \label{substitution} \eea
where the imaginary part $\tilde{\bk}'$ takes the simple form  $\tilde{k}'_j=-\gamma_j/v_j$ for small $\tilde{\bk}$ in this model. Accordingly, we define a ``non-Bloch Hamiltonian'' as follows:  \bea \tilde{H}(\tilde{\bk}) \equiv H(\bk\rw \tilde{\bk} + i\tilde{\bk}'). \label{nonB} \eea  In our model, the replacement $k_j\rw\tilde{k}_j-i\gamma_j/v_j$ leads to
\bea \tilde{H}(\tilde{\bk})= v_x \tilde{k}_x \sigma_x+ v_y \tilde{k}_y \sigma_y  +(\tilde{m} +\frac{t_x\tilde{k}_x^2+t_y\tilde{k}_y^2}{2} -i\sum_j \frac{t_j\gamma_j\tilde{k}_j}{v_j}  )\sigma_z, \quad\quad \label{effective} \eea where \bea \tilde{m}= m -t_x-t_y-\frac{t_x\gamma_x^2}{2v_x^2}-\frac{t_y\gamma_y^2}{2v_y^2}. \eea
The above approach towards $\tilde{H}(\tilde{\bk})$ is quite general and can in principle be implemented directly on lattice models without taking continuum limit. Eq.(\ref{substitution}) and Eq.(\ref{nonB}) remain applicable, though $\tilde{\bk}'$ in general should be treated as a function of $\tilde{\bk}$, which parametrizes a generalized Brillouin zone $\tilde{T}^2(\tilde{\bk})$. It is a deformation of the standard Brillouin zone $T^2(\bk)$ into complex spaces.

Our non-Bloch Chern number is defined as the standard Chern number of $\tilde{H}(\tilde{\bk})$ (not of $H(\bk)$\cite{shen2017topological,esaki2011}).  Because $\tilde{H}(\tilde{\bk})$ is generally non-Hermitian, we define the standard right/left eigenvectors by
\bea  \tilde{H}(\tilde{\bk})\ket{u_{\text{R}\alpha}}=E_\alpha  \ket{u_{\text{R}\alpha}},\quad \tilde{H}^\dag(\tilde{\bk}) \ket{u_{\text{L}\alpha}}=E_\alpha^* \ket{u_{\text{L}\alpha}}, \eea where $\alpha$ is the band index. The normalization $\la u_{\text{L}\alpha}\ket{u_{\text{R}\alpha}}=1$ is required in defining Chern numbers. If we diagonalize $\tilde{H}(\tilde{\bk})= VJ V^{-1}$, $J$ being diagonal, then every column of $V$ (or $(V^\dag)^{-1}$) is a right (or left) eigenvector, with the normalization $\la u_{\text{L}\alpha}\ket{u_{\text{R}\beta}} =\delta_{\alpha\beta}$ satisfied. Now we introduce the non-Bloch Chern number in the generalized Brillouin zone $\tilde{T}^2(\tilde{\bk})$:
\bea C_{(\alpha)}=\frac{1}{2\pi i}\int_{\tilde{T}^2}d^2\tilde{\bk}\,\epsilon^{ij} \bra{\partial_i u_{\text{L}\alpha}(\tilde{\bk})}\partial_j u_{\text{R}\alpha}(\tilde{\bk}) \ra, \label{chern} \eea where $\epsilon^{xy}=-\epsilon^{yx}=1$. Eq.(\ref{chern}) determines the chiral edge modes of open-boundary systems (squares, disks, triangles, etc). It can also be expressed as $C_{(\alpha)} = \frac{1}{2\pi i}\int_{\tilde{T}^2} d^2\tilde{\bk}\, \epsilon^{ij}\text{Tr}(P_\alpha\partial_iP_\alpha\partial_jP_\alpha)$, where the projection operator
$P_\alpha(\tilde{\bk})=\ket{u_{\text{R}\alpha}(\tilde{\bk})}\bra{u_{\text{L}\alpha}(\tilde{\bk})}$.

For the present two-band model, we shall focus on the Chern number of the ``valence band'' ($\text{Re}(E_\alpha)<0$), omitting the $\alpha$ index in Eq.(\ref{chern}).  We compute the Chern number from Eq.(\ref{effective}), and obtain that $C=1$ (0) for $\tilde{m}<0$ ($>0$). When $t_{x,y}=v_{x,y}=1,\,\gamma_{x,y}=\gamma$, the topologically-nontrivial condition $\tilde{m}<0$ becomes $m<2+\gamma^2$, and the phase boundary is
\bea m=2+\gamma^2,  \label{quadratic} \eea which is confirmed by our numerical calculations (see Fig.\ref{phase}). We note that in the low-energy theory, $\gamma$ is treated as being small, and we can see from Fig.\ref{phase} that $\gamma\sim 0.5$ remains well described. As a comparison, the Bloch Chern number\cite{shen2017topological,esaki2011} is nonzero only when $m<2-\sqrt{2}\gamma$; moreover, the Bloch Chern number cannot be defined for $m\in [2-\sqrt{2}\gamma,2+\sqrt{2}\gamma]$ because the bands are gapless (inseparable).

To summarize our approach: We calculate the imaginary part  $\tilde{\bk}'$ of wavevector, which is then used to generate $\tilde{H}(\tilde{\bk})$. The non-Bloch Chern number is then defined via $\tilde{H}(\tilde{\bk})$ in a standard manner. The calculation is simplified in the continuum-model approach, which does not require any numerical input. For certain models, we have calculated the non-Bloch Chern number directly from the lattice models\footnote{For example, the non-Bloch Chern number of $H(\bk)=\sum_{j=x,y}\sin (k_j+i\gamma_j)\sigma_j +[m-t\sum_j\cos(k_j+i\gamma_j)]\sigma_z$ jumps from $1$ to $0$ at $m=2t$ (independent of the value of $\gamma_{x,y}$)\cite{supplemental}.}. It will be useful to develop efficient  algorithms to calculate $\tilde{\bk}'$ and $C$ beyond the continuum approach, which is left to future studies.

\begin{figure}
\subfigure{\includegraphics[width=4.2cm, height=3.7cm]{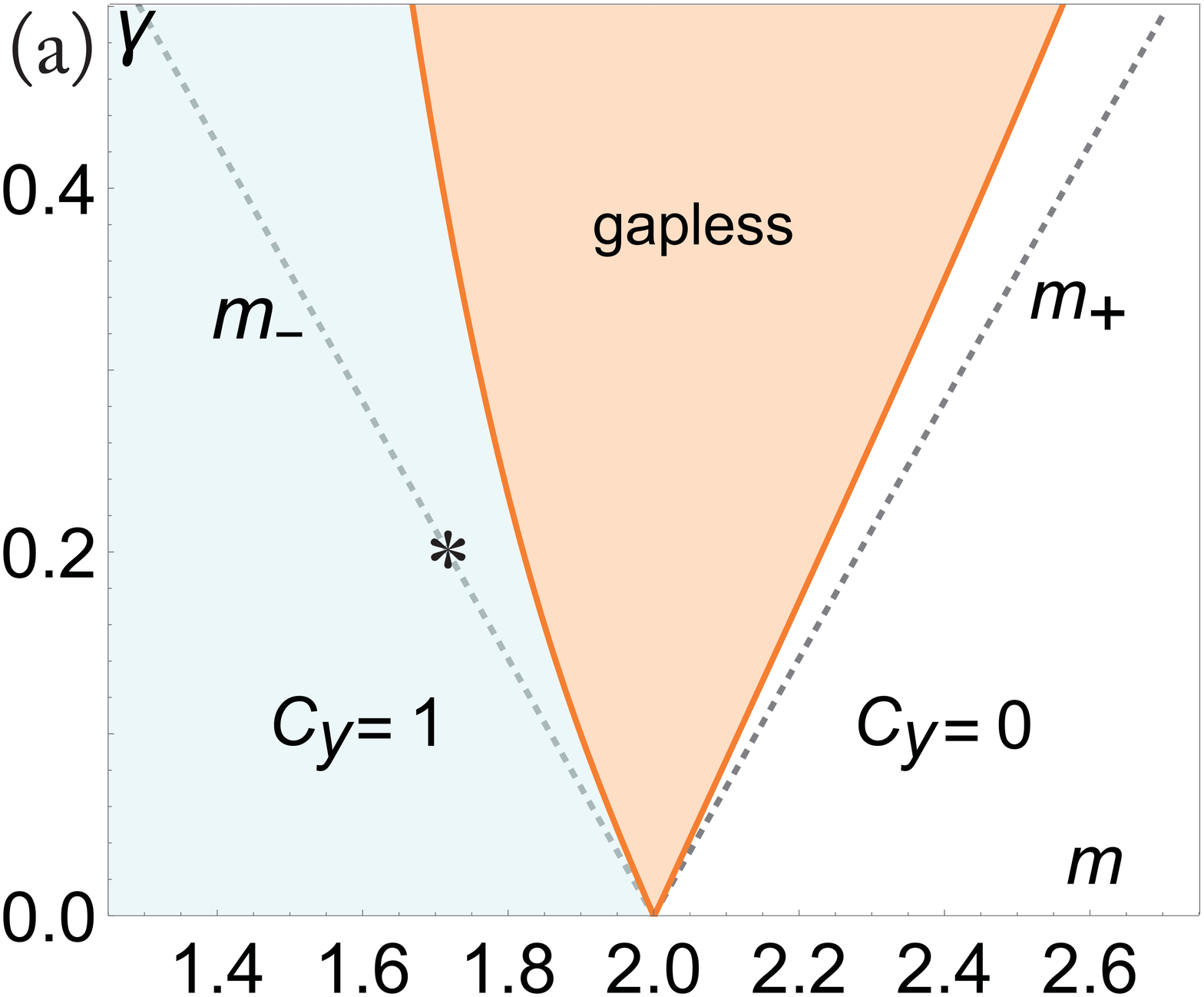}}
\subfigure{\includegraphics[width=4.2cm, height=3.7cm]{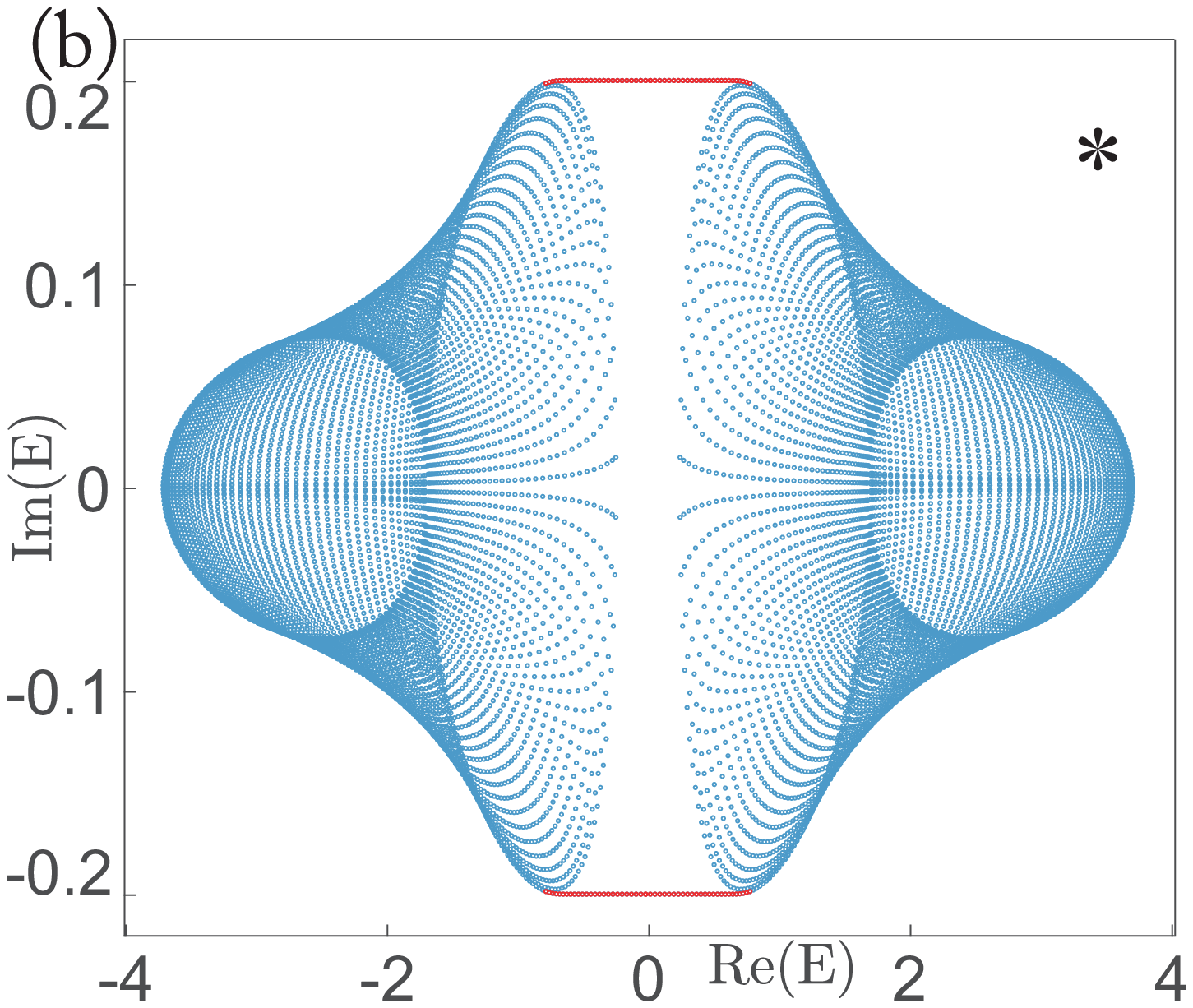}}
\caption{ (a) Phase diagram based on the spectra on a cylinder with open boundary condition in the $y$ direction. $t_{x,y}=v_{x,y}=1,\,\gamma_{x,y}=\gamma,\,\gamma_z=0$. The dotted lines are the Bloch phase boundaries. (b) The spectra for $(m,\gamma)=(1.717,0.2)$ (indicated as $\ast$ in (a)).  The cylinder height is $L_y=40$, and $180$ grid points are taken for $k_x$. The spectra for all $k_x$'s are shown together in the complex plane, without specifying the $k_x$ values. The chiral edge states are shown in red.  }\label{cylinder}
\end{figure}

\emph{Cylinder.--}Now we briefly discuss the cylinder topology whose spectra are noticeably different from the square/disk topology. Suppose that the cylinder has periodic-boundary condition in the $x$ direction and open boundaries in the $y$ direction. The Hamiltonian can be diagonalized as a family of 1D Hamiltonians parametrized by the good quantum number $k_x$. As an illustration, we take a set of parameters indicated as $\ast$ in Fig.\ref{cylinder}(a), and show the numerical spectra in Fig.\ref{cylinder}(b). Topological edge states can be readily seen in the spectra.

In fact, to characterize the chiral edge states on the cylinder, one can define a non-Bloch ``cylinder Chern number'', which is denoted as $C_y$ for the open boundaries in $y$ direction. The definition is quite similar to Eq.(\ref{chern}), except that $(\tilde{k}_x,\tilde{k}_y)$ is replaced by $(k_x,\tilde{k}_y)$, because the eigenstates are forced to be Bloch waves in the $x$ direction. A non-Bloch ``cylinder Hamiltonian'' $\tilde{H}_y(k_x,\tilde{k}_y)$ can be obtained from $H(\bk)$ via $k_y\rw\tilde{k}_y+i\tilde{k}'_y$ (similar to Eq.(\ref{nonB})), then $C_y$ can be defined by $\tilde{H}_y$, which we shall not repeat due to the resemblance to the construction of $C$ [Eq.(\ref{chern})].

We would like to emphasize the following:
(i) The value of non-Bloch cylinder Chern number depends on the edge orientation. For example, if we take open boundaries in the $x$ direction, the Chern number $C_x$ defined by $\tilde{H}_x(\tilde{k}_x,k_y)$ can be different from $C_y$. (ii) The original non-Bloch Chern number defined in Eq.(\ref{chern}) is the physically more useful one. In fact, we find that wave-packet motions on the edges of cylinder follow the phase diagram of  Fig.\ref{phase}, namely, chiral edge motions are appreciable when $C$ (instead of $C_y$) is nonzero. This is understandable because wave packets are quite ignorant of the periodic-boundary condition in the $x$ direction if the cylinder circumference is much larger than the wave packet size.

\emph{Conclusions.--}We uncovered a non-Bloch bulk-boundary correspondence: The chiral edge states are determined by non-Bloch Chern numbers defined in the complex Brillouin zone. The obtained phase diagrams (also confirmed numerically) are qualitatively different from the Bloch-Hamiltonian counterparts. Our results suggest a non-Bloch framework for non-Hermitian band topology.

There are many open questions ahead. For example, it is worthwhile to study the respective roles of the Bloch and non-Bloch Chern numbers: What aspects of non-Hermitian physics are described by the Bloch/non-Bloch one? In addition, the theory can be generalized to many other topological non-Hermitian systems. It is also interesting to go beyond the band theory (e.g., to consider interaction effects).

{\it Acknowledgements.--}We would like to thank Hui Zhai for discussions. This work is supported by NSFC under grant No. 11674189.

\bibliography{dirac}

\vspace{11mm}

{\bf Supplemental Material}

\vspace{4mm}

This is a supplemental material of ``Non-Hermitian Chern bands''. It contains: (i) Technical details in calculating the phase diagram; (ii) Geometry independence of the phase diagram; (iii) Phase diagrams for other values of parameters; (iv) Pictorial illustration of non-Hermitian skin effect; (v) Solution on cylinder topology; (vi) Phase diagram and non-Bloch Chern number of an analytically solvable model.

\section{I. Technical details in the calculation of phase diagram}

\begin{figure}[htb]
{\includegraphics[width=7cm, height=7cm]{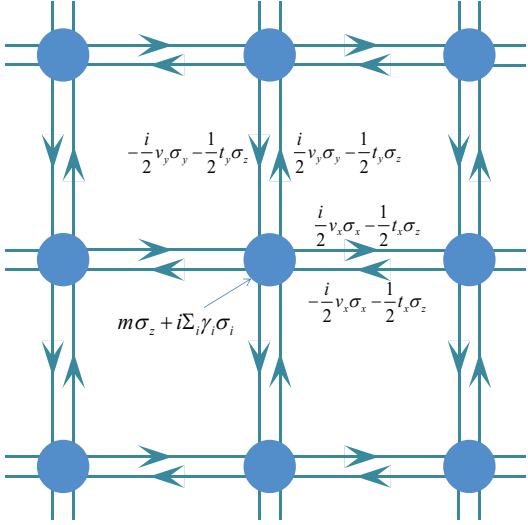}}
\caption{Pictorial illustration of the real-space Hamiltonian. The corresponding Bloch Hamiltonian has been given in Eq.(1) in the main article.   }\label{picH}
\end{figure}

In real space, the Hamiltonian of our model reads
\begin{equation}
\begin{aligned}
\hat{H}=&[\sum_\bx \sum_{j=x,y} c^\dagger_\bx (-\frac{i}{2}v_j\sigma_j-\frac{1}{2}t_j\sigma_z) c_{\bx+\be_j} +H.c]\\
&+\sum_\bx c^\dagger_\bx (m\sigma_z+i\sum_{j=x,y,z}\gamma_j\sigma_j) c_\bx,
\end{aligned} \label{}
\end{equation} where $\bx=(x,y)$ are the integer coordinates of unit cells, $c_\bx=(c_{\bx,A},c_{\bx,B})^T$ is a two-component annihilation operator, and $\be_j$ is the unit vector along the $j$ direction. A pictorial illustration is given in Fig.\ref{picH}.

\begin{figure}[htb]
{\includegraphics[width=5.0cm, height=4.2cm]{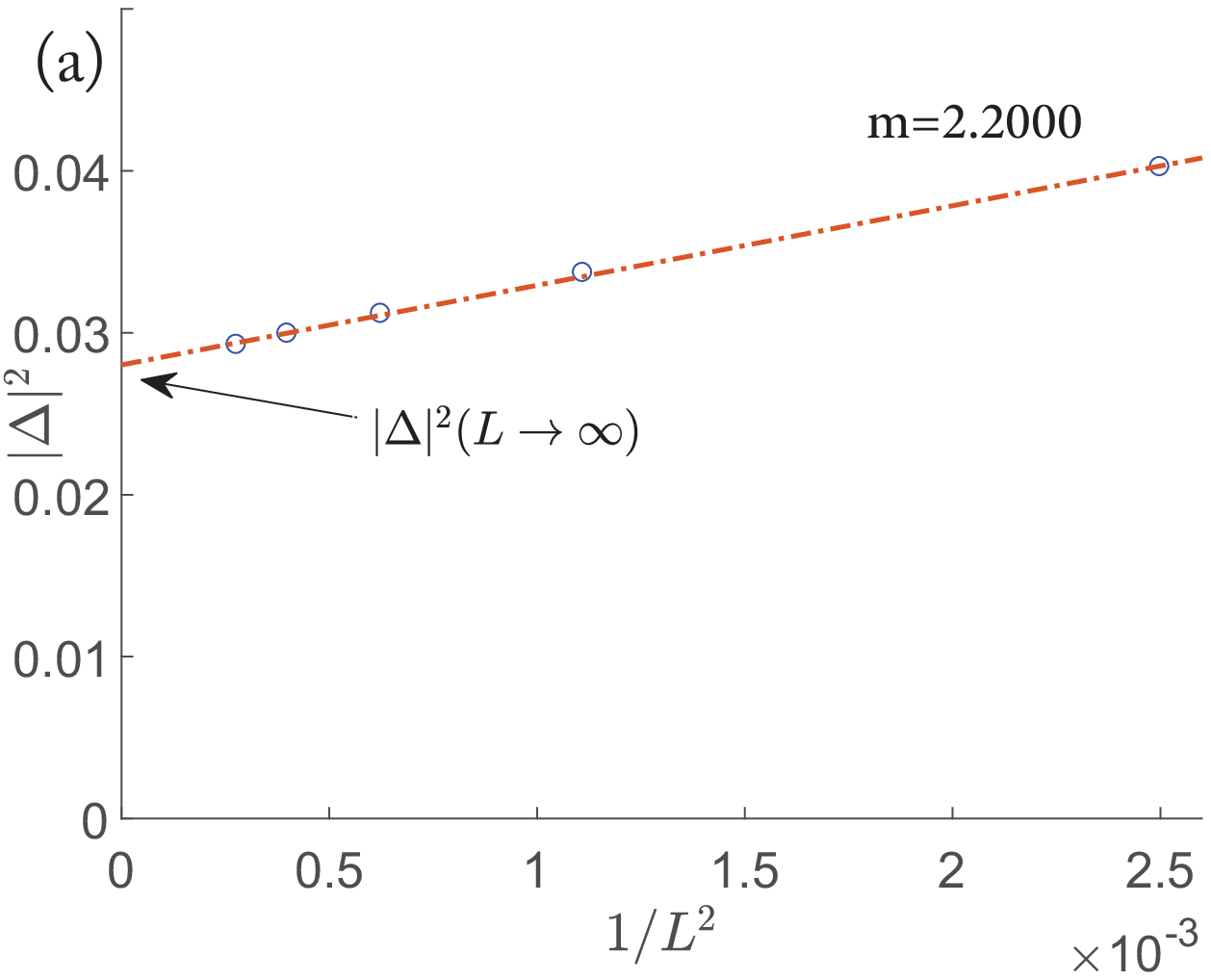}}
{\includegraphics[width=5.0cm, height=4.2cm]{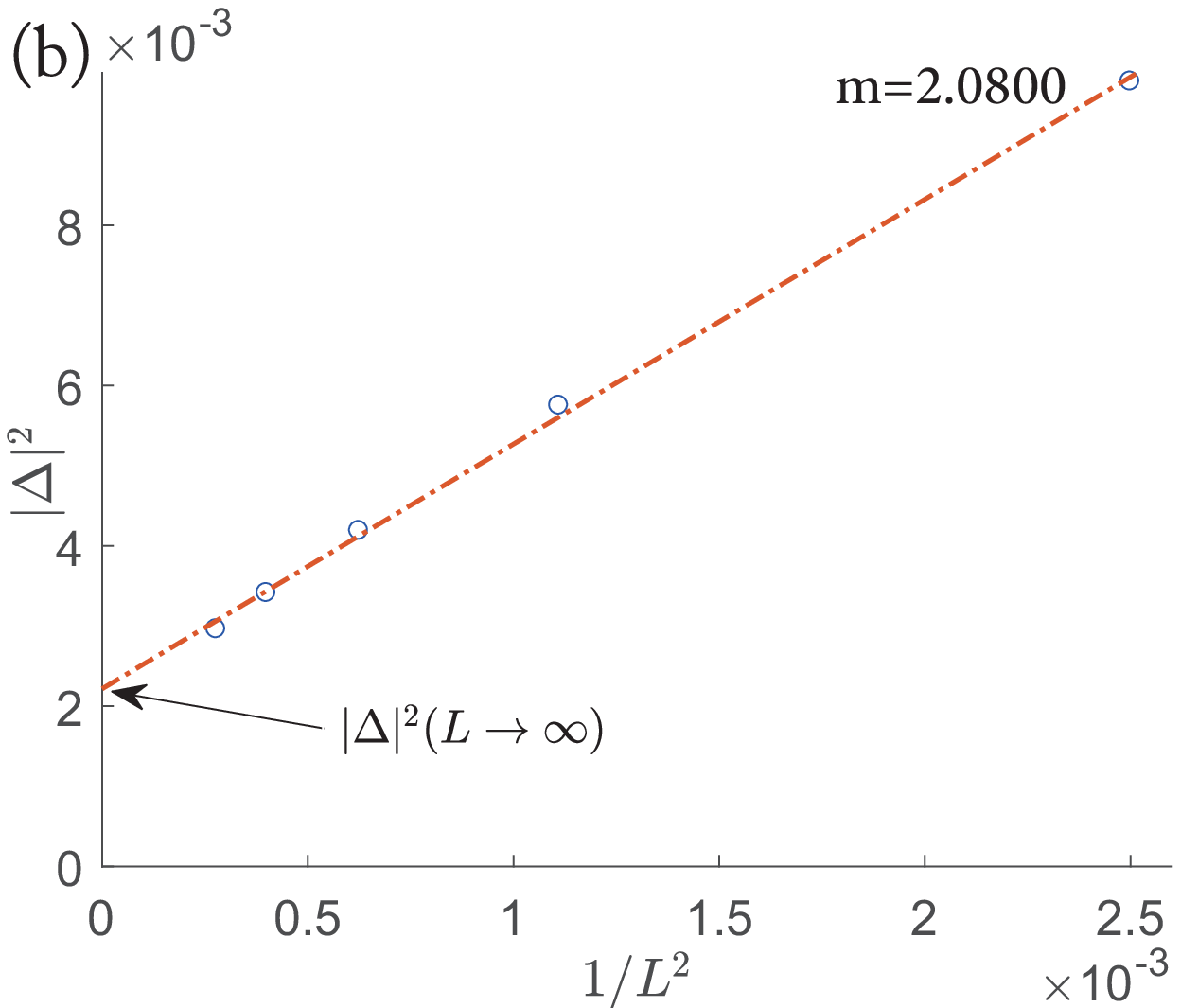}}
{\includegraphics[width=5.0cm, height=4.2cm]{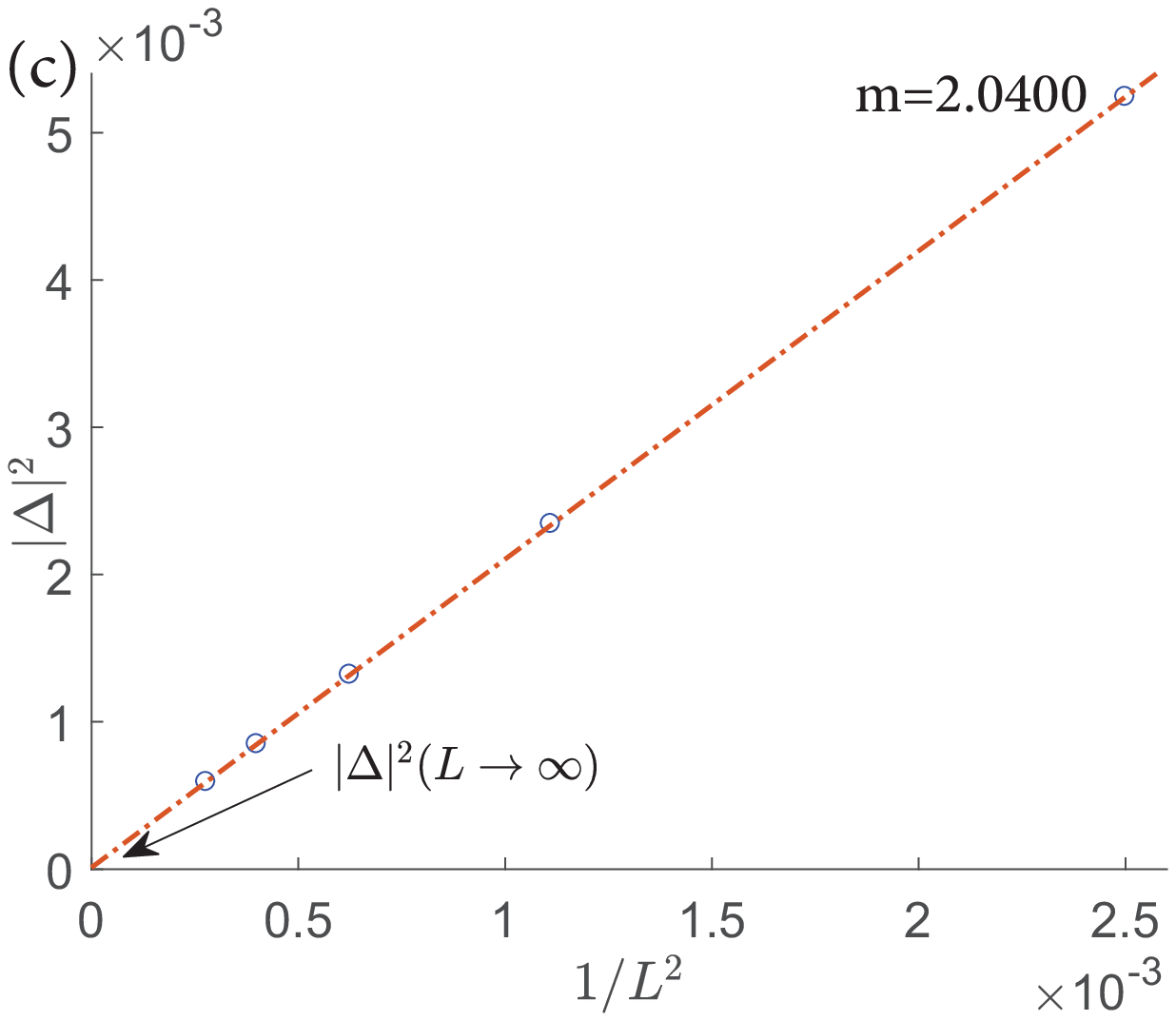}}
\caption{ The magnitude of the gap square $|\Delta|^2$ as a function of $1/L^2$, where $L$ is the radius of disk (We take disk geometry in this illustration; the method is the same for square geometry). (a) $m=2.2000$, (b) $m=2.0800$, (c) $m=2.0400$.
$t_x=t_y=1$,$v_x=v_y=1$,$\gamma_x=\gamma_y=0.2$. The intercept of $|\Delta|^2$-$1/L^2$ line gives the gap square in the $L\rw\infty$ limit. For (a) and (b), the gaps are nonzero; while for (c) the gap vanishes.     }\label{fitting}
\end{figure}

We have obtained open-boundary spectra and phase diagram by solving the real-space Hamiltonian on various geometries including squares (i.e., $1\leq x,y\leq L$) and disks (i.e., $x^2+y^2\leq L^2$) with varying size $L$. To obtain an accurate phase boundary, we plot the gap square $|\Delta|^2$ as a function of $1/L^2$. The intercept of linear fitting gives the gap square $|\Delta|^2$ in the $L\rw\infty$ limit. Several examples of fitting are shown in Fig.\ref{fitting} for the disk geometry. The phase boundary is obtained by finding the gap-closing points, namely $|\Delta|^2(L\rw\infty)=0$. These calculations lead to the phase diagram.

\section{II. Geometry independence of the phase diagram}

In the main article, the phase diagram (Fig.1) obtained from square geometry has been shown. To establish the topological nature of the phase diagram, we should show that it is independent of the geometry of the system. To this end, we have done calculations on other geometries including disks (i.e., $x^2+y^2\leq L^2$) and triangles, and have indeed found the same phase diagram.

\begin{figure}[htb]
{\includegraphics[width=8cm, height=5.0cm]{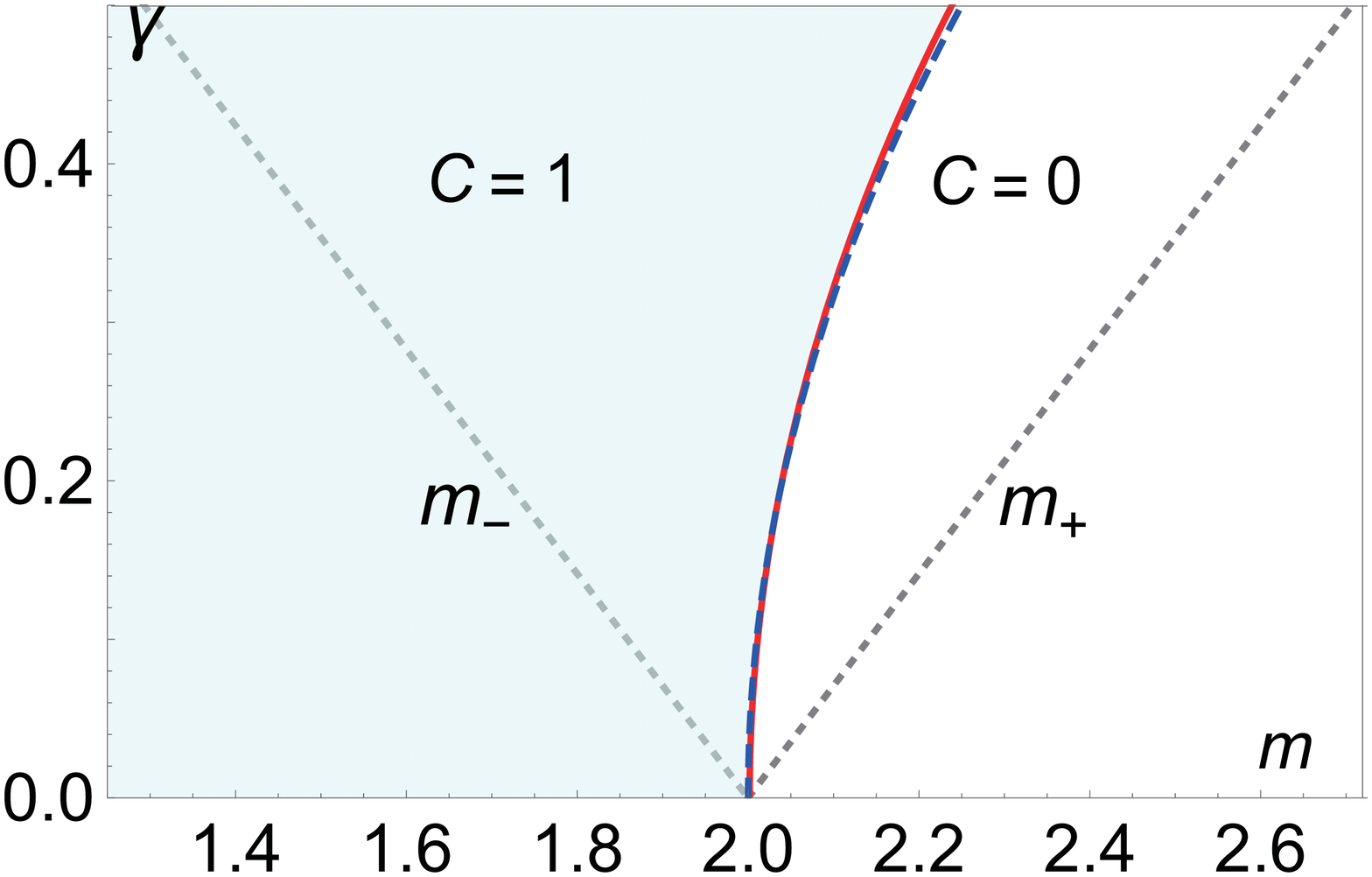}}
\caption{ Phase diagram obtained from disk geometry. The red solid curve is the numerical phase boundary, while the dashed curve (very close to the solid curve) is the phase boundary obtained from the non-Bloch theory ($m=2+\gamma^2$). The dotted lines $m_\pm$ are the Bloch phase boundary. This phase diagram is indistinguishable from the one obtained from square geometry (Fig.1 in the main article). This insensitivity to the geometrical shapes is a manifestation of the topological nature of the physics. }\label{diskphase}
\end{figure}

The phase diagram from the disk geometry is shown in Fig.\ref{diskphase}, which is indistinguishable from the one obtained from the square geometry (Fig.1 in the main article). This insensitivity to geometrical shapes enables a topological characterization.

\begin{table}[htbp]
  \centering
  \caption{The numerical data used in plotting Fig.\ref{diskphase}. Here, ``$m_c$ numerical'' stands for the gap-closing point as $m$ is tuned, and ``$m_c$ theory'' is the value obtained from the non-Bloch theory (using the continuum model, see the main article).  The red solid curve in Fig.\ref{diskphase} is based on ``$m_c$ numerical'' (with error $<3\times 10^{-4}$), while the dashed curve is based on ``$m_c$ theory'' ($m_c=2+\gamma^2$).  }
     \begin{tabular}{ |c|c|c|}
\hline
            $\gamma$     &  $m_c$ numerical   & $m_c$ theory ($2+\gamma^2$)    \\
\hline
           0.05     & 2.0025     & 2.0025      \\
\hline
           0.10     & 2.0100     & 2.0100      \\
\hline
           0.15     & 2.0225     & 2.0225      \\
\hline
           0.20     & 2.0400     & 2.0400      \\
\hline
           0.25     & 2.0625     & 2.0625      \\
\hline
          0.30     & 2.0885     & 2.0900      \\
\hline
           0.35     & 2.1200     & 2.1225      \\
\hline
           0.40     & 2.1540     & 2.1600      \\
 \hline
           0.45     & 2.1940     & 2.2025      \\
 \hline
           0.50     & 2.2360     & 2.2500      \\
\hline
    \end{tabular}%
  \label{diskdata}%
\end{table}%

\section{III. Phase diagrams for other choices of parameters}

In the main article, we calculated the phase diagram for $t_x=t_y=v_x=v_y=1$, $\gamma_x=\gamma_y=\gamma$. Here, we provide phase diagrams for a few other values of parameter.

\begin{figure}[htb]
{\includegraphics[width=8cm, height=5.0cm]{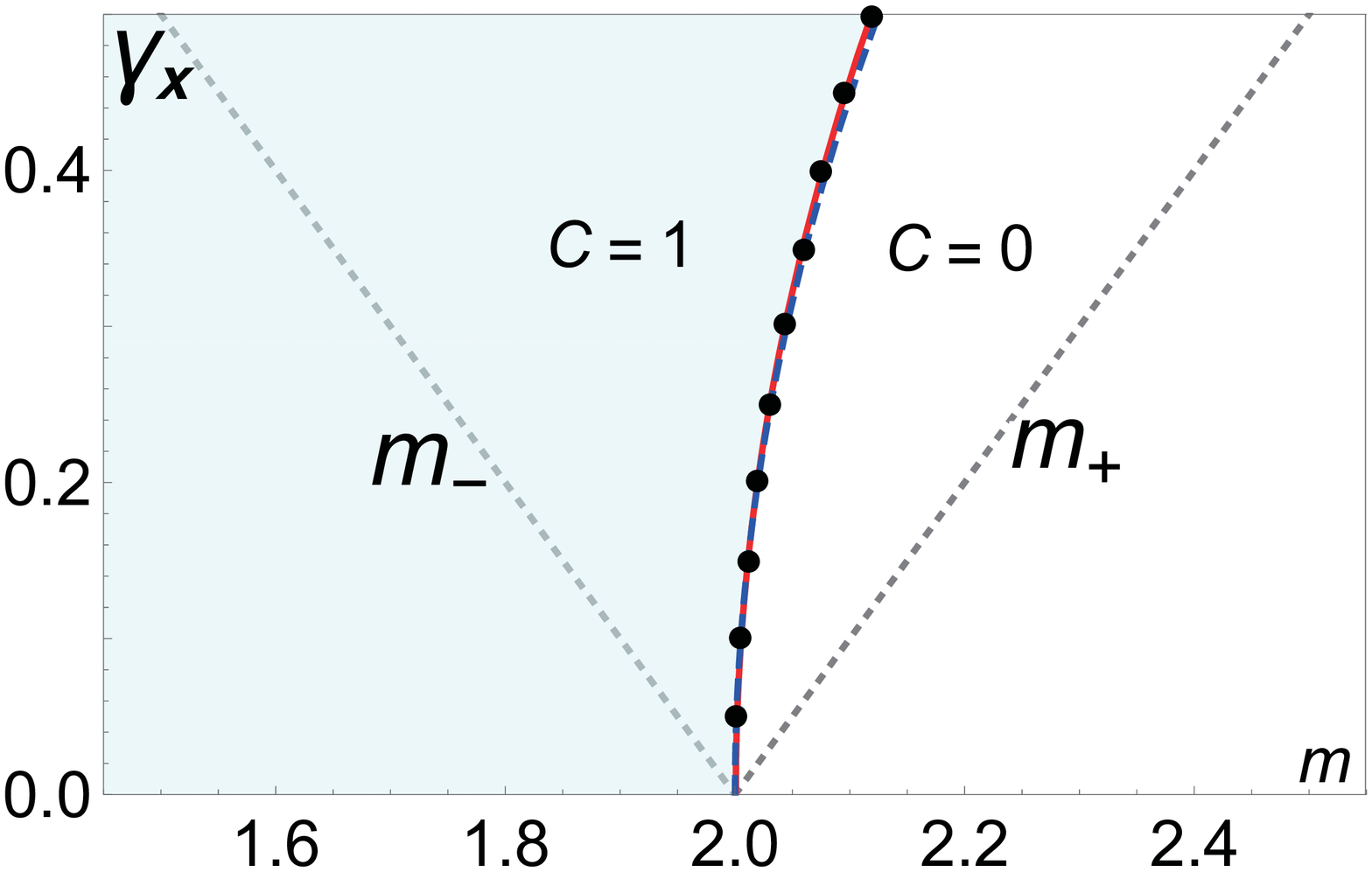}}
{\includegraphics[width=8cm, height=5.0cm]{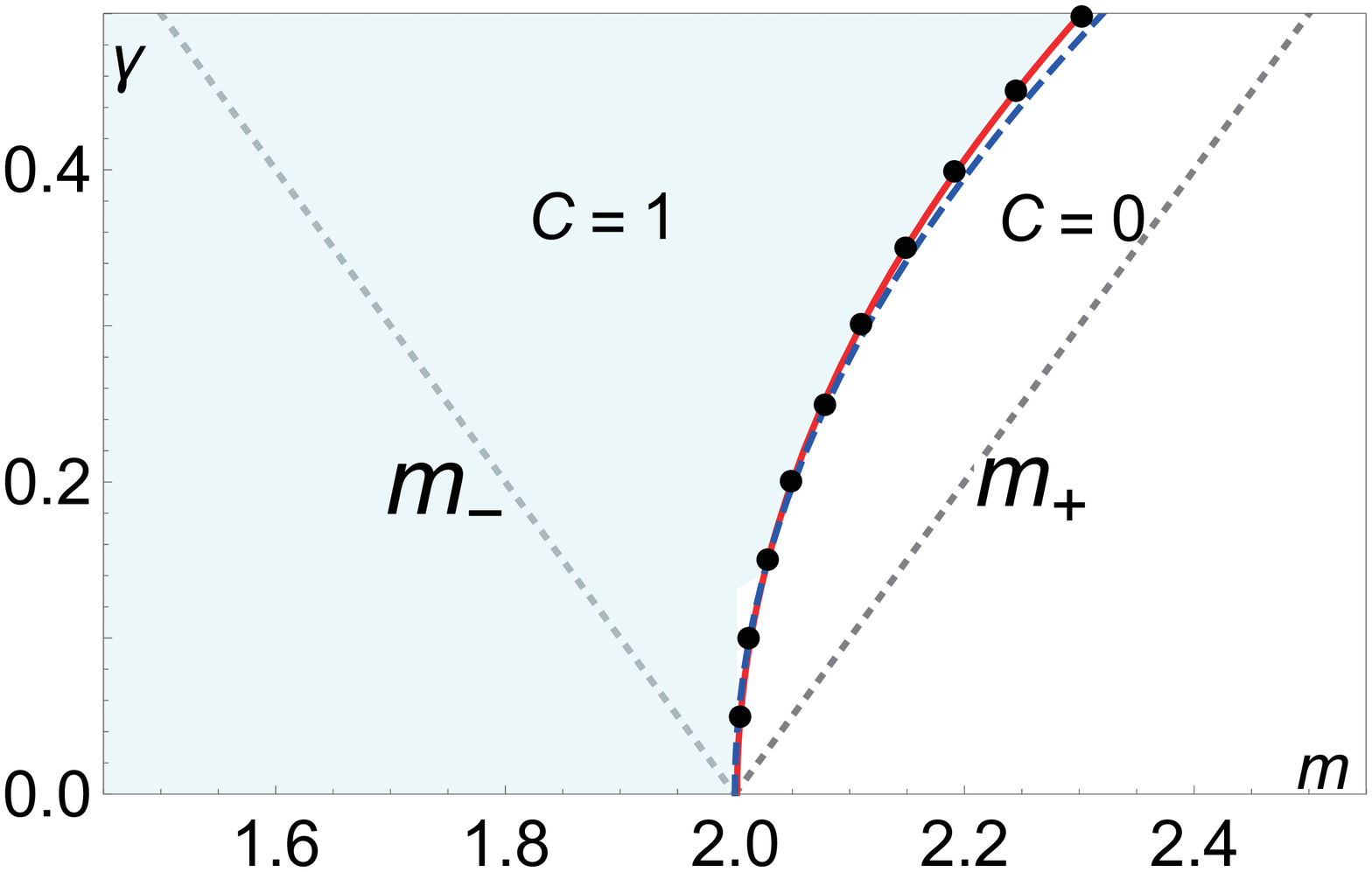}}
\caption{ Upper panel: Phase diagram for $\gamma_x\neq 0, \gamma_y=0$. Other parameters are $t_x=t_y=1$,$v_x=v_y=1$. The red solid curve is the numerical phase boundary, while the blue dashed curve (very close to the red solid curve) is the theoretical prediction of non-Bloch theory: $m=2+\gamma_x^2/2$. Lower panel: Phase diagram for $t_x=t_y=1$, $v_x=0.8,v_y=1$, and $\gamma_x=\gamma_y=\gamma$. The dashed curve is the prediction of (low-energy) non-Bloch theory:  $m=2+1.28125\gamma^2$. }\label{gammax}
\end{figure}

The upper panel of Fig.\ref{gammax} is the phase diagram for $\gamma_x\neq 0$ and $\gamma_y=0$. Now the theoretically predicted phase boundary is $m=2+\gamma_x^2/2$, which is in agreement with the numerical result.

The lower panel of Fig.\ref{gammax} is for $t_x=t_y=v_y=1$, $v_x=0.8$, and $\gamma_x=\gamma_y=\gamma$. The agreement between theory and numerical results is confirmed again. Note that the theoretical prediction of phase boundary, namely $m=2+1.28125\gamma^2$, is based on the low-energy continuum model whose $\gamma$ is treated as being small; therefore, slight deviation from the numerical results for large $\gamma$ is natural.

%\begin{table}[htbp]
%  \centering
%  \caption{Numerical data used in plotting the phase diagram in Fig.\ref{gammax} (Upper panel). }
%    \begin{tabular}{ |c|c|c|}
%\hline
%           $\gamma_x$     & $m_c$ numerical    & $m_c$ theory ($2+\gamma_x^2/2$)     \\
%\hline
%           0.1     & 2.005     & 2.005      \\
%\hline
%           0.2     & 2.020     & 2.020      \\
%\hline
%           0.3     & 2.044     & 2.045      \\
%\hline
%           0.4     & 2.077     & 2.080      \\
%\hline
%    \end{tabular}%
%  \label{}%
%\end{table}

\section{IV. Illustration of non-Hermitian skin effect}

\begin{figure}[htb]
{\includegraphics[width=4.2cm, height=3.8cm]{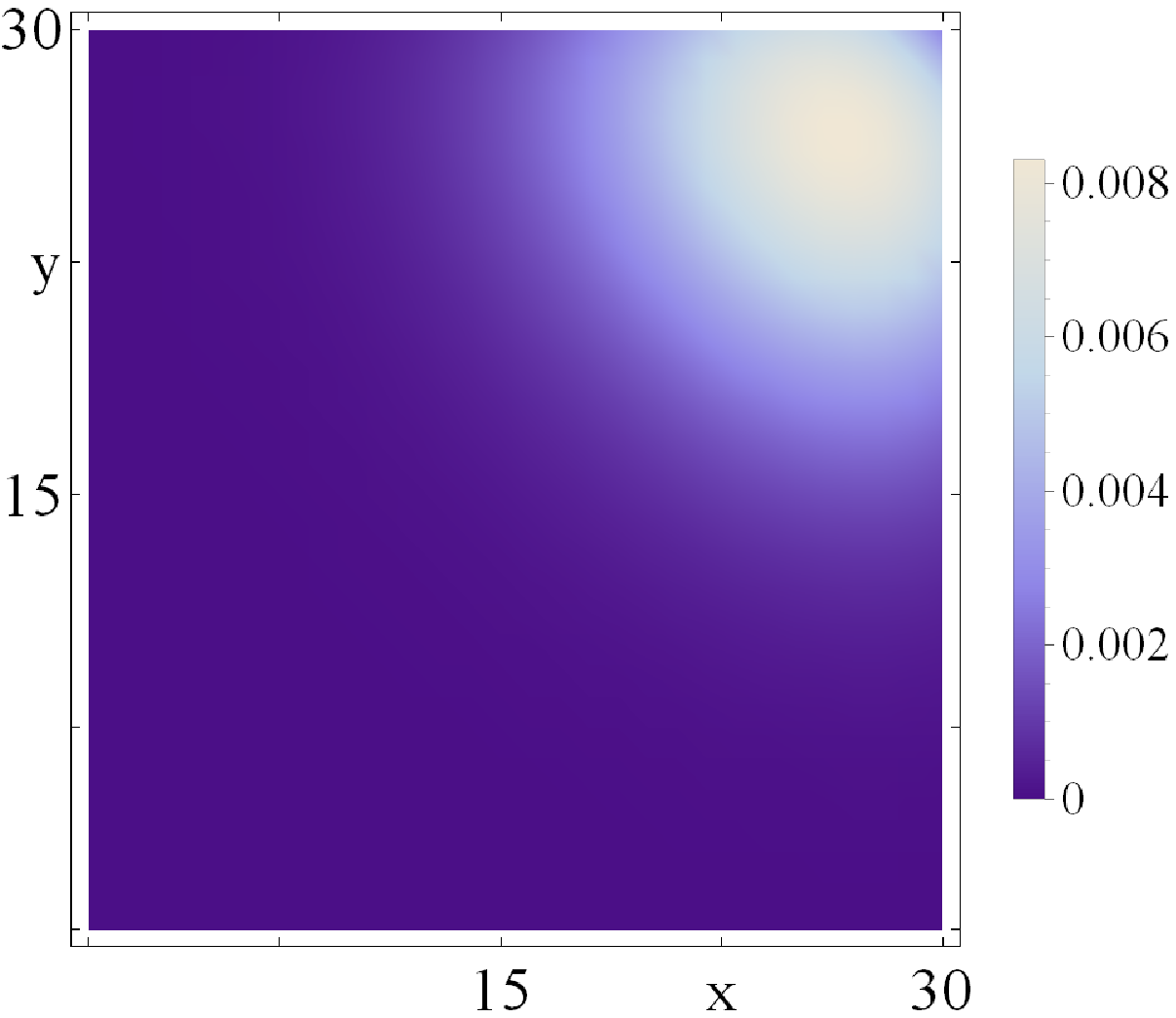}}
{\includegraphics[width=4.2cm, height=3.8cm]{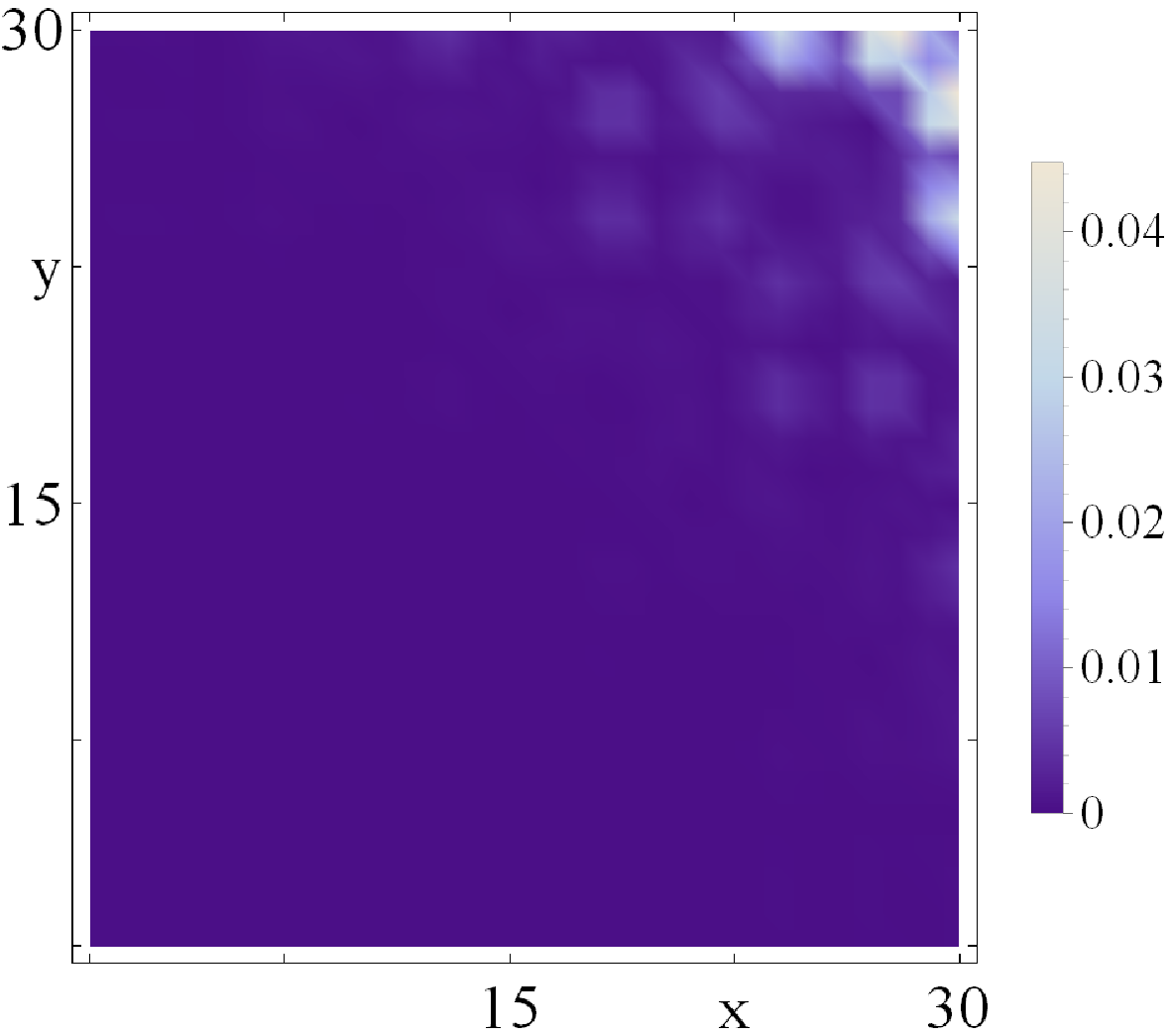}}
{\includegraphics[width=4.2cm, height=3.7cm]{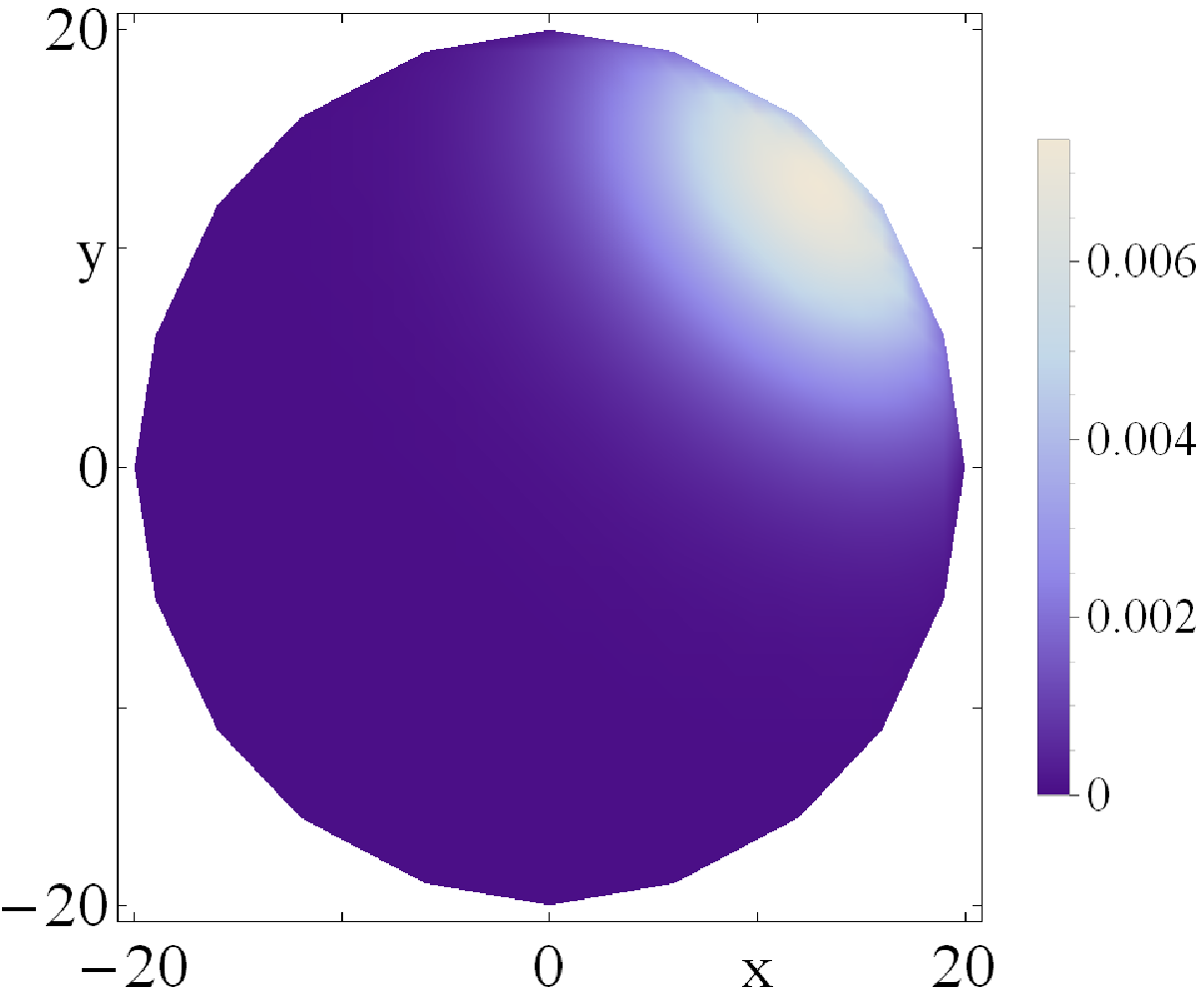}}
{\includegraphics[width=4.2cm, height=3.7cm]{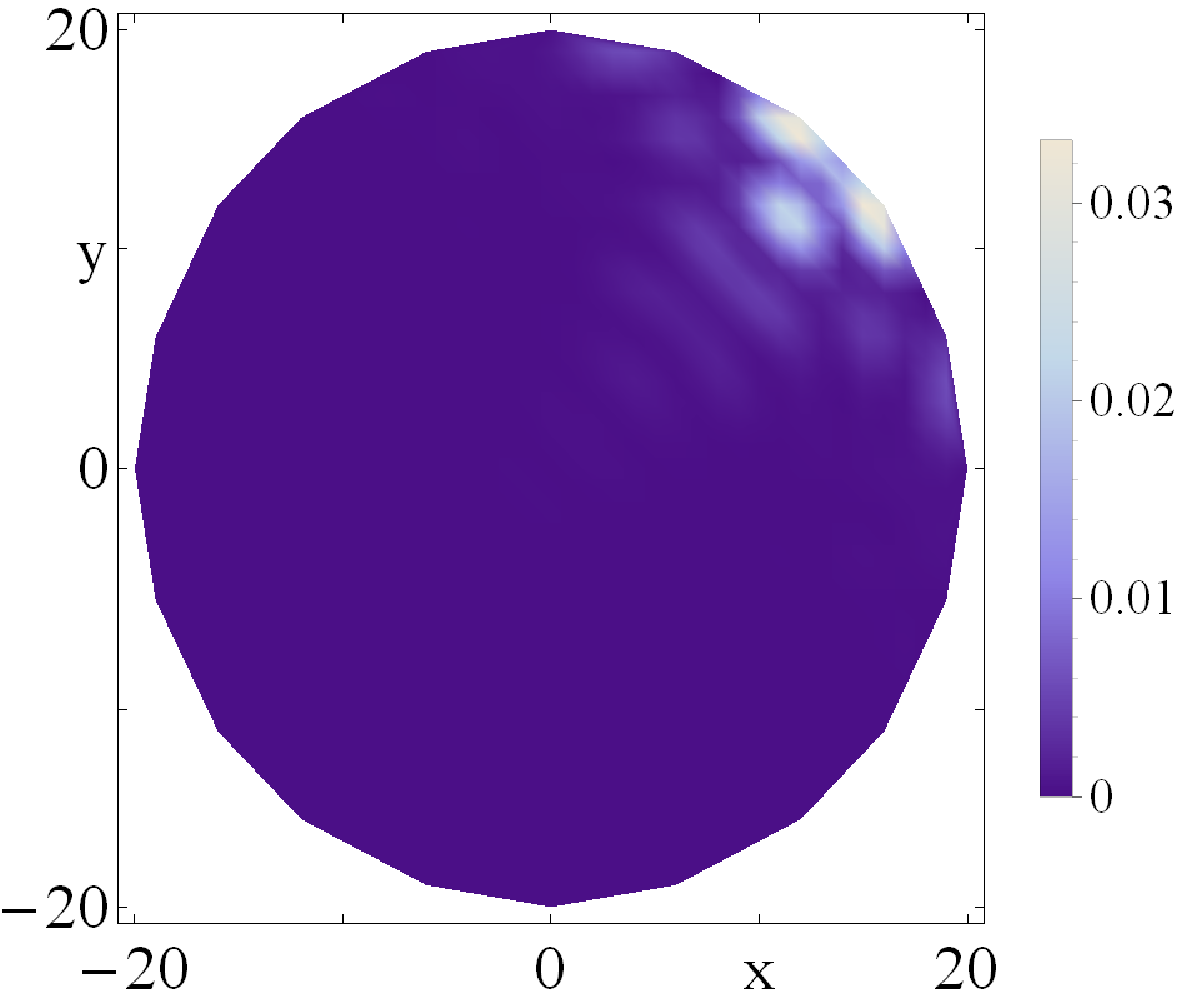}}
\caption{ Typical profiles of bulk eigenstates on square and disk. Parameters: $t_x=t_y=1$,$v_x=v_y=1$, $\gamma_x=\gamma_y=0.15$, $m=2.2121$.  }\label{skin1}
\end{figure}

\begin{figure}[htb]
{\includegraphics[width=4.2cm, height=3.8cm]{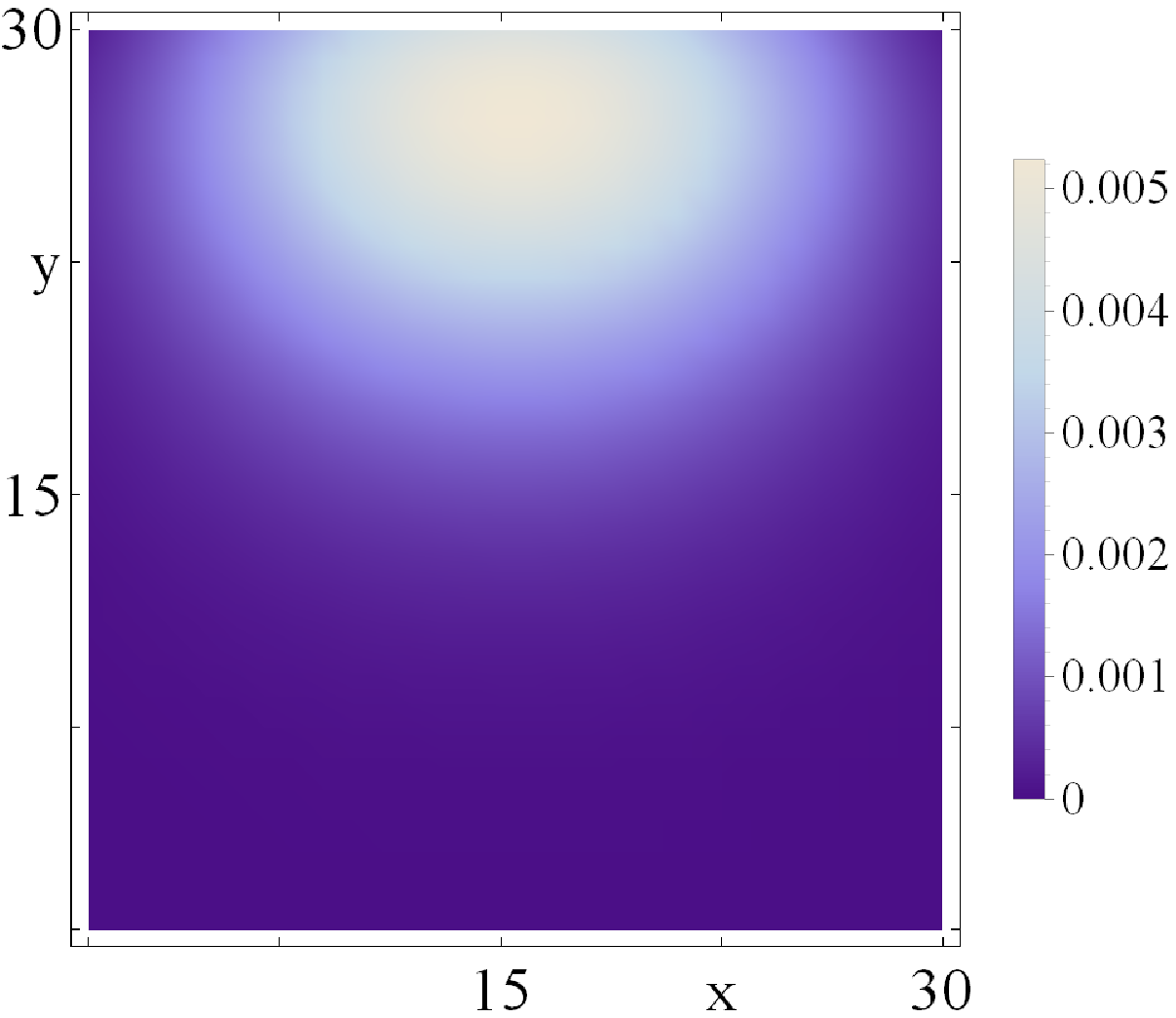}}
{\includegraphics[width=4.2cm, height=3.8cm]{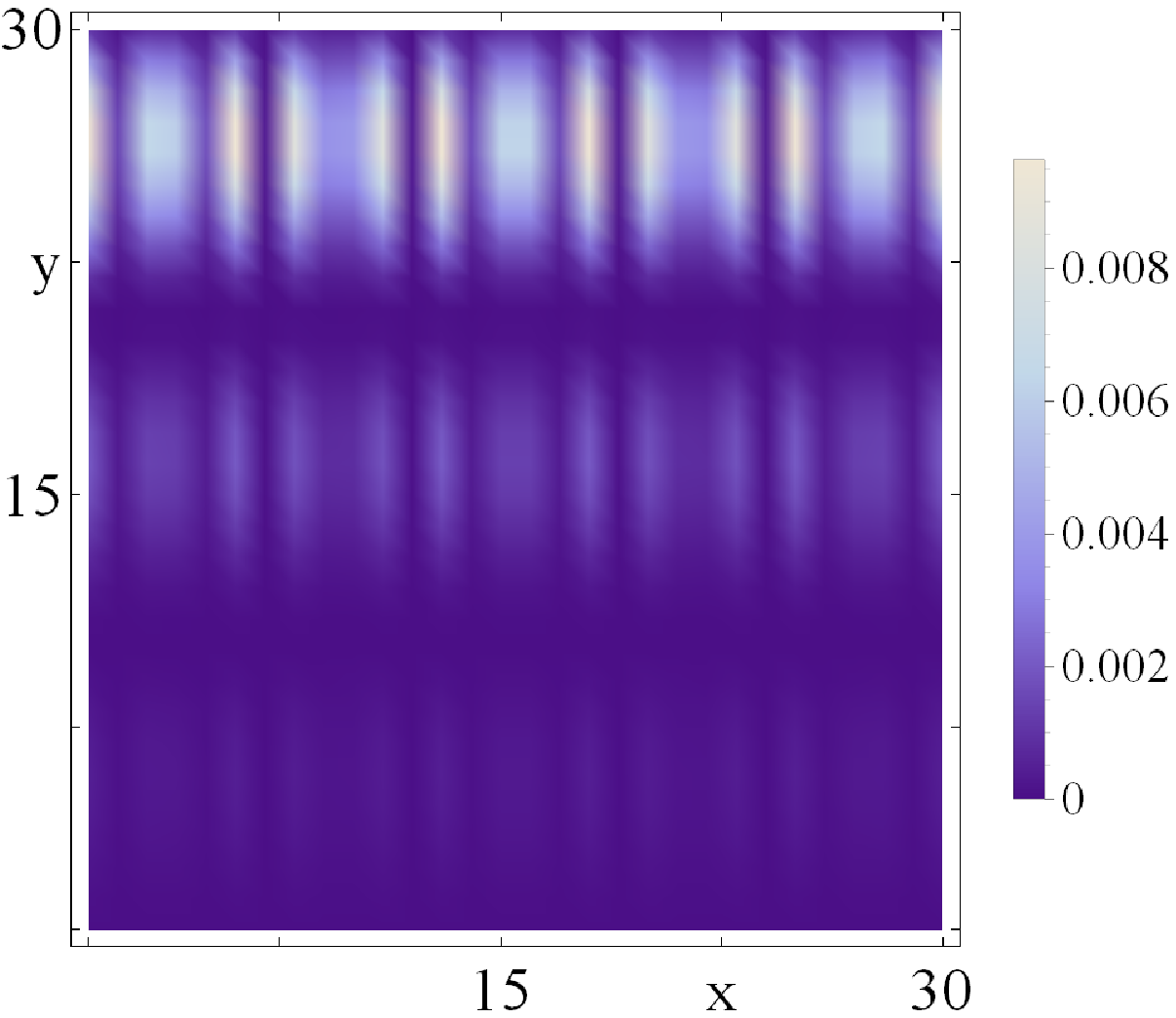}}
{\includegraphics[width=4.2cm, height=3.7cm]{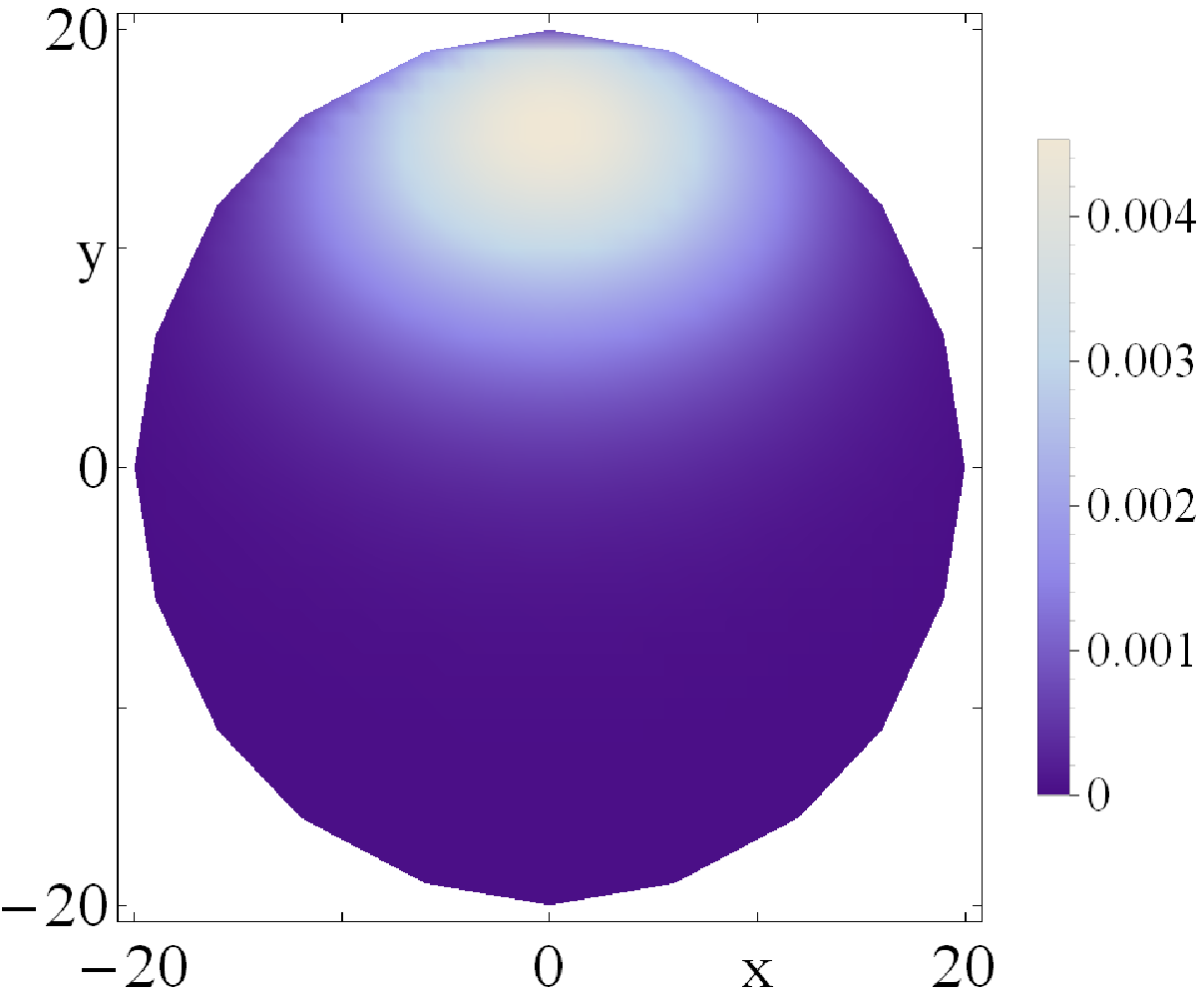}}
{\includegraphics[width=4.2cm, height=3.7cm]{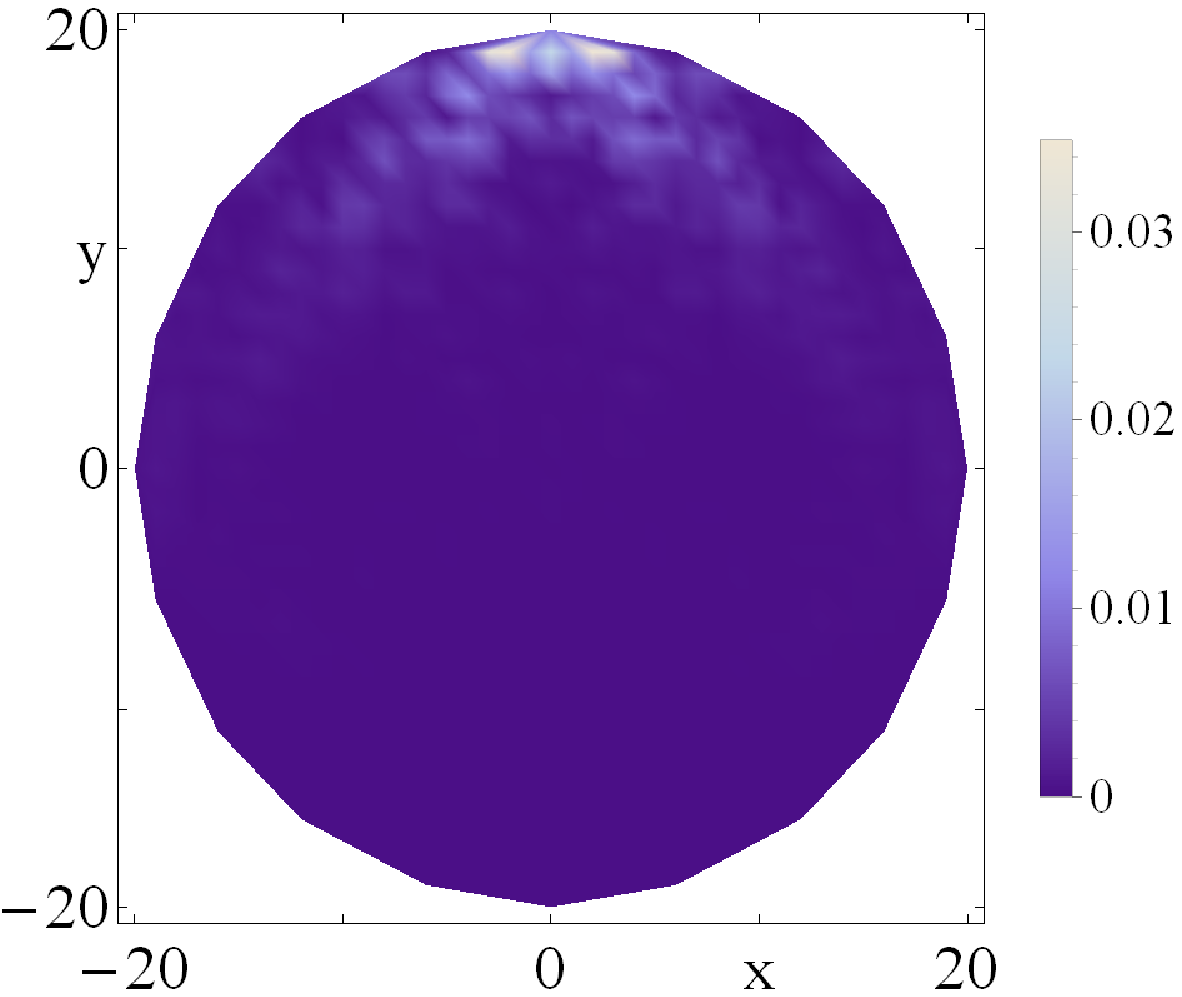}}
\caption{Typical profiles of bulk eigenstates on square and disk, for $t_x=t_y=1$, $v_x=v_y=1$, $\gamma_x=0$, $\gamma_y=0.15$, $m=2.2121$ (The same as Fig.\ref{skin1} except that $\gamma_x=0$).   }\label{skin2}
\end{figure}

As pictorial illustrations of the non-Hermitian skin effect,
we show profiles of several typical bulk eigenstates (i.e., eigenstates in the continuum spectrum). In Fig.\ref{skin1} we have $\gamma_x=\gamma_y=0.15$, while in Fig.\ref{skin2},  $\gamma_x=0,\gamma_y=0.15$. All eigenstates are exponentially localized at the boundary of systems (i.e., non-Hermitian skin effect).

\section{V. Solution of cylinder topology}

In the main article, we have briefly discussed the cylinder topology. Here, we provide more details. It is more convenient to do a basis transformation of Eq.(1) of the main article: $(\sigma_x,\sigma_y,\sigma_z)\rw(\sigma_z,-\sigma_y,\sigma_x)$. We take $t_i=v_i$, then Eq.(1) becomes
\begin{equation}
\begin{aligned}
H(\bk)=(t_x\sin k_x+i\gamma_x)\sigma_z- (t_y\sin k_y+i\gamma_y)\sigma_y&\\+(m -t_x\cos k_x -t_y\cos k_y)\sigma_x,
\end{aligned} \label{cylinderH}
\end{equation}
Let us consider a cylinder with open-boundary condition in the $y$ direction with height $L$. Now $k_x$ remains a good quantum number, and the wavefunction can be written as a $2L$-component vector $\ket{\psi(k_x)}=(\psi_{1,A},\psi_{1,B}, \cdots,\psi_{L,A},\psi_{L,B})^T$ with $k_x$ implicit. The eigenvalue equation $H\ket{\psi}=E\ket{\psi}$ gives the bulk equations
$-t_y\psi_{n-1,B}+(t_x\sin k_x+i\gamma_x)\psi_{n,A}+(m -t_x\cos k_x-\gamma_y)\psi_{n,B}=E\psi_{n,A}$, and
$(m -t_x\cos k_x+\gamma_y)\psi_{n,A}-(t_x\sin k_x+i\gamma_x)\psi_{n,B}-t_y\psi_{n+1,A}=E\psi_{n,B}$.
These bulk equations can be satisfied by taking
\bea\label{ansatz} (\psi_{n,A},\psi_{n,B}) = \beta^{n}(\phi_A,\phi_B), \eea
where $\beta$ and $\phi_{A,B}$ satisfy
\begin{equation} \label{bulkeigen}
\begin{aligned}
&[(m-t_x\cos k_x-\gamma_y)-t_y\beta^{-1}]\phi_{B}   =(E-t_x\sin k_x-i\gamma_x)\phi_{A},\\
&[(m-t_x\cos k_x+\gamma_y)-t_y\beta]\phi_{A}   =(E+t_x\sin k_x+i\gamma_x)\phi_{B}.
\end{aligned}
\end{equation} It follows that
\bea \label{bulkeigen2}
\begin{aligned}
&[(m-t_x\cos k_x-\gamma_y)-t_y\beta^{-1}][(m-t_x\cos k_x+\gamma_y)-t_y\beta]\\&=E^2-(t_x\sin k_x+i\gamma_x)^2,
\end{aligned}
\eea
There are multiple solutions $\beta_i$'s and $(\phi^{(i)}_{n,A},\phi^{(i)}_{n,B})$ to Eq.(\ref{bulkeigen}), and the boundary conditions require that the true eigenstates are superpositions of these solutions, namely, $(\psi_{n,A},\psi_{n,B})=\sum_i \beta_i^n(\phi^{(i)}_{A},\phi^{(i)}_{B})$. In this model, for an eigen-energy $E$, we have two solutions of $\beta$ from the quadratic Eq.(\ref{bulkeigen2}). They are denoted as $\beta_{1,2}(E)$. The open-boundary condition in the $y$ direction requires that $|\beta_1(E)|=|\beta_2(E)|$ if $E$ belongs to the continuum spectrum.  The derivation of this statement is similar to Ref.\cite{yao2018edge}. In fact, the boundary equations\footnote{A boundary condition equivalent to Eq.(\ref{boundarycond}) is to add two auxiliary wavefunction $\psi_{0,B}$ and $\psi_{L+1,A}$ with $\psi_{0,B}=\psi_{L+1,A}=0$, which also leads to Eq.(\ref{equalmod}).}
\begin{equation}
\begin{aligned}
&(m -t_x\cos k_x-\gamma_y)\psi_{1,B}=(E -t_x\sin k_x-i\gamma_x )\psi_{1,A}\\
&(m-t_x\cos k_x+\gamma_y)\psi_{L,A}=(E + t_x\sin k_x+i\gamma_x))\psi_{L,B},
\end{aligned}\label{boundarycond}
\end{equation}
requires that
\begin{equation}
\begin{aligned}
&[(m -t_x\cos k_x-\gamma_y)-t_y\beta_1^{-1}]\beta_1^{L+1}\\
&=[(m -t_x\cos k_x-\gamma_y)-t_y\beta_2^{-1}]\beta_2^{L+1}.
\end{aligned}  \label{equalmod}
\end{equation} Now we can see that the continuum spectrum necessitates $|\beta_1(E)|=|\beta_2(E)|$. Otherwise, suppose that $|\beta_1(E)|<|\beta_2(E)|$, then the left hand side of Eq.(\ref{equalmod}) is negligible when $L$ is large, and Eq.(\ref{equalmod}) becomes $(m -t_x\cos k_x-\gamma_y)-t_y\beta_2^{-1}(E)=0$ or $\beta_2(E)=0$. This is inconsistent with the fact that the number of energy eigenstates in the continuum spectrum is proportional to $L$. This argument is similar to Ref.\cite{yao2018edge}.

With $|\beta_1|=|\beta_2|$ as input, Vieta's formula applied to Eq.(\ref{bulkeigen2}) tells us that
\begin{equation} \label{bulkbetay}
\begin{aligned}
|\beta_{1,2}| = \sqrt{|\frac{m-t_x\cos k_x+\gamma_y}{m-t_x\cos k_x-\gamma_y}|} \equiv r(k_x).
\end{aligned}
\end{equation} As explained in the main article (the ``cylinder'' section), in the non-Bloch theory we should use complex-valued $k_y$: \bea k_y\rw \tilde{k}_y+i\tilde{k}'_y, \label{replaceHy} \eea while $k_x$ remains real-valued (see the main article). According to the calculations above, we have  \bea \tilde{k}'_y= -\ln|\beta_{1,2}| =-\ln r(k_x). \eea
Inserting Eq.(\ref{replaceHy}) into Eq.(\ref{cylinderH}), we have the non-Bloch ``cylinder Hamiltonian'':
\bea \tilde{H}_y(k_x,\tilde{k}_y) &=& H(k_x\rw k_x,k_y\rw\tilde{k}_y-i\ln r) \nn \\ &=& (t_x\sin k_x +i\gamma_x)\sigma_z-i\gamma_y\sigma_y + (m-t_x\cos k_x)\sigma_x \nn \\ && -t_y r^{-1}e^{-i\tilde{k}_y}\sigma_+ -t_y re^{i\tilde{k}_y}\sigma_-, \eea
in which $r=\sqrt{|\frac{m-t_x\cos k_x+\gamma_y}{m-t_x\cos k_x-\gamma_y}|}$ and $\sigma_\pm=(\sigma_x\pm i\sigma_y)/2$. The cylinder ``phase diagram''  can be plotted using the energy spectra of $\tilde{H}_y(k_x,\tilde{k}_y)$.  Alternatively, we can calculate the non-Bloch cylinder Chern number $C_y$, which is defined as the usual Chern number of $\tilde{H}_y(k_x,\tilde{k}_y)$ (see the main article). The value of $C_y$ is numerically found to be $1$ and $0$ in the two gapped regions (see Fig.3 of the main article), respectively, and $C_y$ strongly fluctuates in the gapless region. In this way, the phase diagram is completely determined by calculating $C_y$.

%In the original basis:
%\begin{equation}
%\begin{aligned}
%\tilde{H}_{y}&=[t_x\sin(k_x)+i\gamma_x]\sigma_x+(\frac{it_y |\beta|^{-1}e^{-ik_y}-it_y|\beta|e^{ik_y}}{2}+i\gamma_y)\sigma_y\\&+[m-t_x\cos(k_x)-\frac{t_y |\beta|^{-1}e^{-ik_y}+t_y|\beta|e^{ik_y}}{2}]\sigma_z,
%\end{aligned} \label{}
%\end{equation}

\section{VI. An analytically solvable model}

\begin{figure}[htb]
{\includegraphics[width=6.2cm, height=6.2cm]{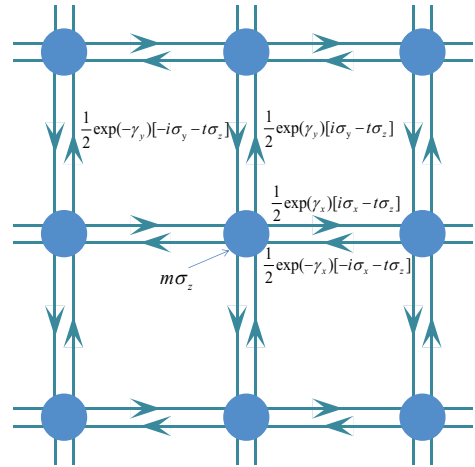}}
\caption{ Pictorial illustration of the model in Eq.(\ref{exactly}). }\label{real2}
\end{figure}

\begin{figure}[htb]
{\includegraphics[width=8.0cm, height=5.0cm]{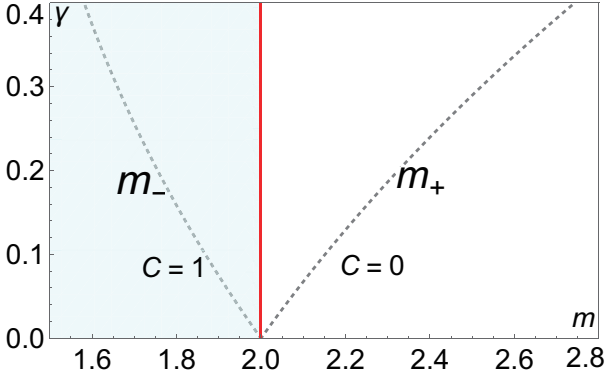}}
\caption{The phase diagram of the analytical solvable model [Eq.(\ref{exactlybloch} or Eq.(\ref{exactly})]. $t=1$, $\gamma_x=\gamma_y=\gamma$. The phase boundary $m=2$ is independent of the values of $\gamma_{x,y}$. The dotted lines $m_\pm$ are the phase boundaries of the Bloch Hamiltonian $H(\bk)$.   }\label{phaseexact}
\end{figure}

In this section we study a two-band model whose non-Bloch Chern number can be calculated directly from the lattice model (without taking the continuum limit). This model is somewhat less nontrivial than Eq.(1) of the main article, yet the dramatic difference between Bloch and non-Bloch theory is also appreciable. The Bloch Hamiltonian reads
\begin{equation}
\begin{aligned}
H(\bk)&=\sin(k_x+i\gamma_x)\sigma_x+\sin(k_y+i\gamma_y)\sigma_y\\&+[m-t\cos(k_x+i\gamma_x)-t\cos(k_y+i\gamma_y)]\sigma_z.
\end{aligned} \label{exactlybloch}
\end{equation}
The real-space Hamiltonian is (see Fig.\ref{real2}):
\begin{equation}
\begin{aligned}
\hat{H}=&\sum_\bx c^\dagger_\bx (m\sigma_z) c_\bx +\sum_\bx c^\dagger_\bx \sum_{j=x,y} \frac{\exp(-\gamma_j)}{2}(-i\sigma_j-t\sigma_z) c_{\bx+\be_j}\\
&+\sum_\bx \sum_{j=x,y} c^\dagger_\bx \frac{\exp(\gamma_j)}{2}(i\sigma_j-t\sigma_z) c_{\bx-\be_j},
\end{aligned} \label{exactly}
\end{equation}
where $\bx=(x,y)$ are the integer coordinates of unit cells, $c_\bx=(c_{\bx,A},c_{\bx,B})^T$ is a two-component annihilation operator, and $\be_j$ is the unit vector along the $j$ direction.

This Hamiltonian can be readily solved for an open-boundary system (e.g., square, disk).  For a square with area $L^2$, the real-space Hamiltonian $H$ is a $2L^2\times 2L^2$ matrix. The eigenvalue equation is $H\ket{\psi}=E\ket{\psi}$, in which $\ket{\psi}$ is a $2L^2$-component vector. This eigenvalue equation is equivalent to $\bar{H}\ket{\bar{\psi}}=E\ket{\bar{\psi}}$ with $\ket{\bar{\psi}}=S^{-1}\ket{\psi}$ and $\bar{H}=S^{-1}HS$. We take $S$ to be a diagonal matrix whose diagonal elements are
\bea S(x,y)=\exp(\gamma_x x+\gamma_y y), \eea
then $\gamma_{x,y}$ are absent in $\bar{H}$. In fact, the resultant $\bk$-space Hamiltonian of $\bar{H}$ is just the Qi-Wu-Zhang model\cite{qi2005} with $\bar{H}(\bk)=\sin(k_x)\sigma_x+\sin(k_y)\sigma_y +[m-t\cos(k_x)-t\cos(k_y)]\sigma_z$. The open-boundary energies of the original Hamiltonian Eq.(\ref{exactlybloch}) is exactly the same as those of $\bar{H}$.

Now we calculate the non-Bloch Chern number.
As $\bar{H}$ is Hermitian, its bulk eigenstates $\ket{\bar{\psi}}$ are extended Bloch waves. Therefore, the eigenstates $\ket{\psi}=S\ket{\bar{\psi}}$ of $H$ has exponential decay factor $S(x,y)=\exp(\gamma_x x+\gamma_y y)$. As explained in the main article, we should take for open-boundary eigenstates a complex-valued wavevector \bea \bk\rw \tilde{\bk}+i\tilde{\bk}', \eea
where the imaginary part is  \bea \tilde{\bk}'=(-\gamma_x,-\gamma_y). \eea
The non-Bloch Hamiltonian is therefore
\bea
&&\tilde{H}(\tilde{\bk})  = H(\bk\rw \tilde{\bk}+i\tilde{\bk}')\nn\\ &&=\sin(\tilde{k}_x)\sigma_x+   \sin(\tilde{k}_y)\sigma_y +[m-t\cos(\tilde{k}_x)-t\cos(\tilde{k}_y)]\sigma_z.\quad
\eea  The non-Bloch Chern number of $H(\bk)$ is just the usual Chern number of $\tilde{H}(\tilde{\bk})$, whose expression is (see the main article):
\bea C_{(\alpha)}=\frac{1}{2\pi i}\int_{\tilde{T}^2}d^2\tilde{\bk}\,\epsilon^{ij} \bra{\partial_i u_{\text{L}\alpha}(\tilde{\bk})}\partial_j u_{\text{R}\alpha}(\tilde{\bk}) \ra, \eea In the present model $\tilde{H}(\tilde{\bk})$ is Hermitian, therefore $\ket{u_{\text{R}\alpha}(\tilde{\bk})}=\ket{u_{\text{L}\alpha}(\tilde{\bk})}$. Let us focus on the $E_\alpha<0$ band and omit the band index $\alpha$. This non-Bloch Chern number is found to be $C=0$ for $m>2t$ and $C=1$ for $0<m<2t$ (taking $t>0$). The phase boundary \bea m=2t \eea is independent of the value of $\gamma_{x,y}$. Nevertheless, it should be emphasized that even in this somewhat boring example, the phase diagram differs from that of Bloch Hamiltonian (see Fig.\ref{phaseexact}), whose phase boundary is \bea m=t(\cosh\gamma_x+\cosh\gamma_y) \pm\sqrt{\sinh^2\gamma_x+\sinh^2\gamma_y}, \eea which is indicated as $m_\pm$ in Fig.\ref{phaseexact}.  The chiral edge modes is tied to the non-Bloch Chern number instead of the Bloch Chern numbers\cite{shen2017topological,esaki2011} of $H(\bk)$.

\begin{figure}[htb]
{\includegraphics[width=4.2cm, height=3.5cm]{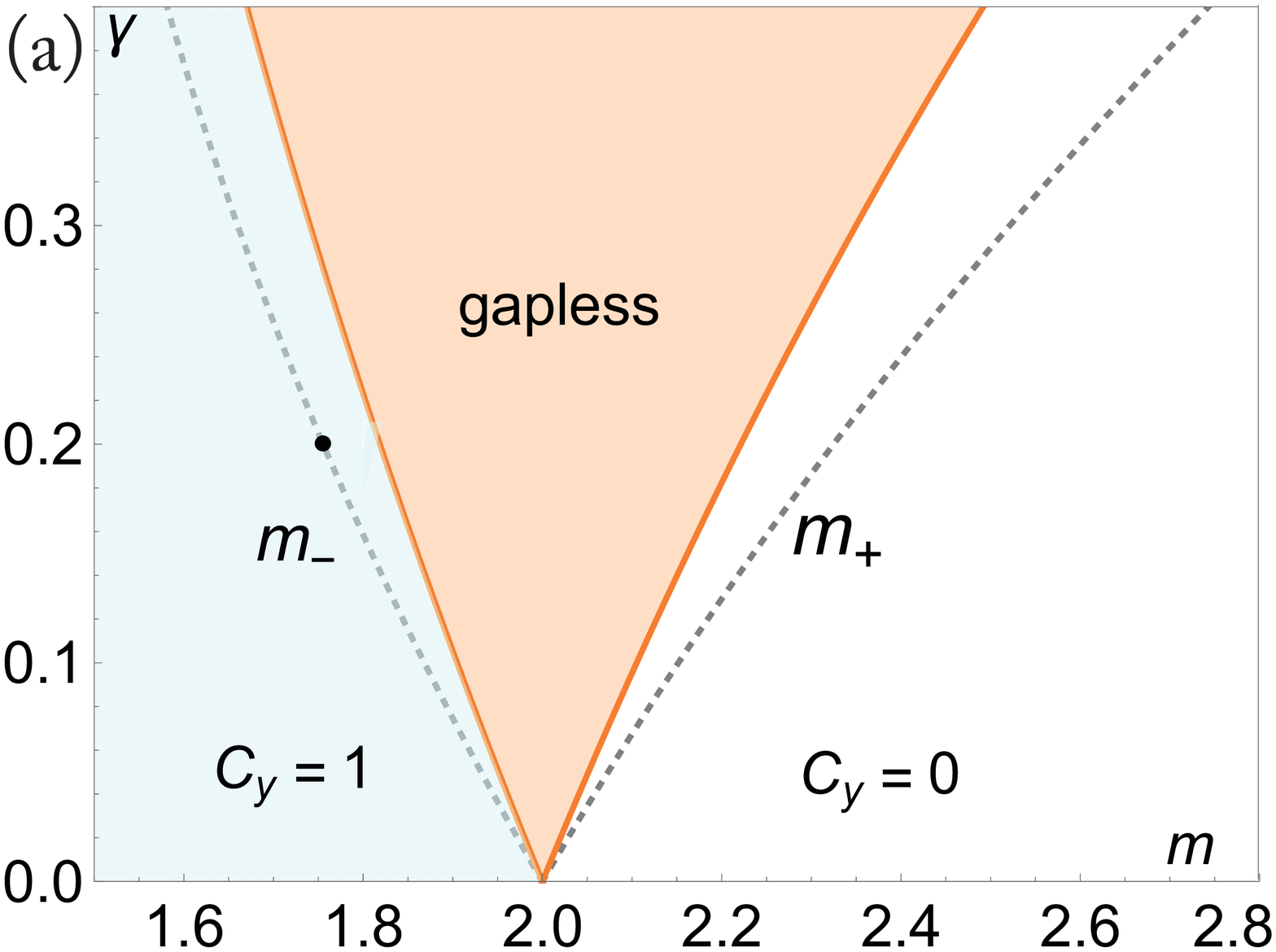}}
{\includegraphics[width=4.2cm, height=3.5cm]{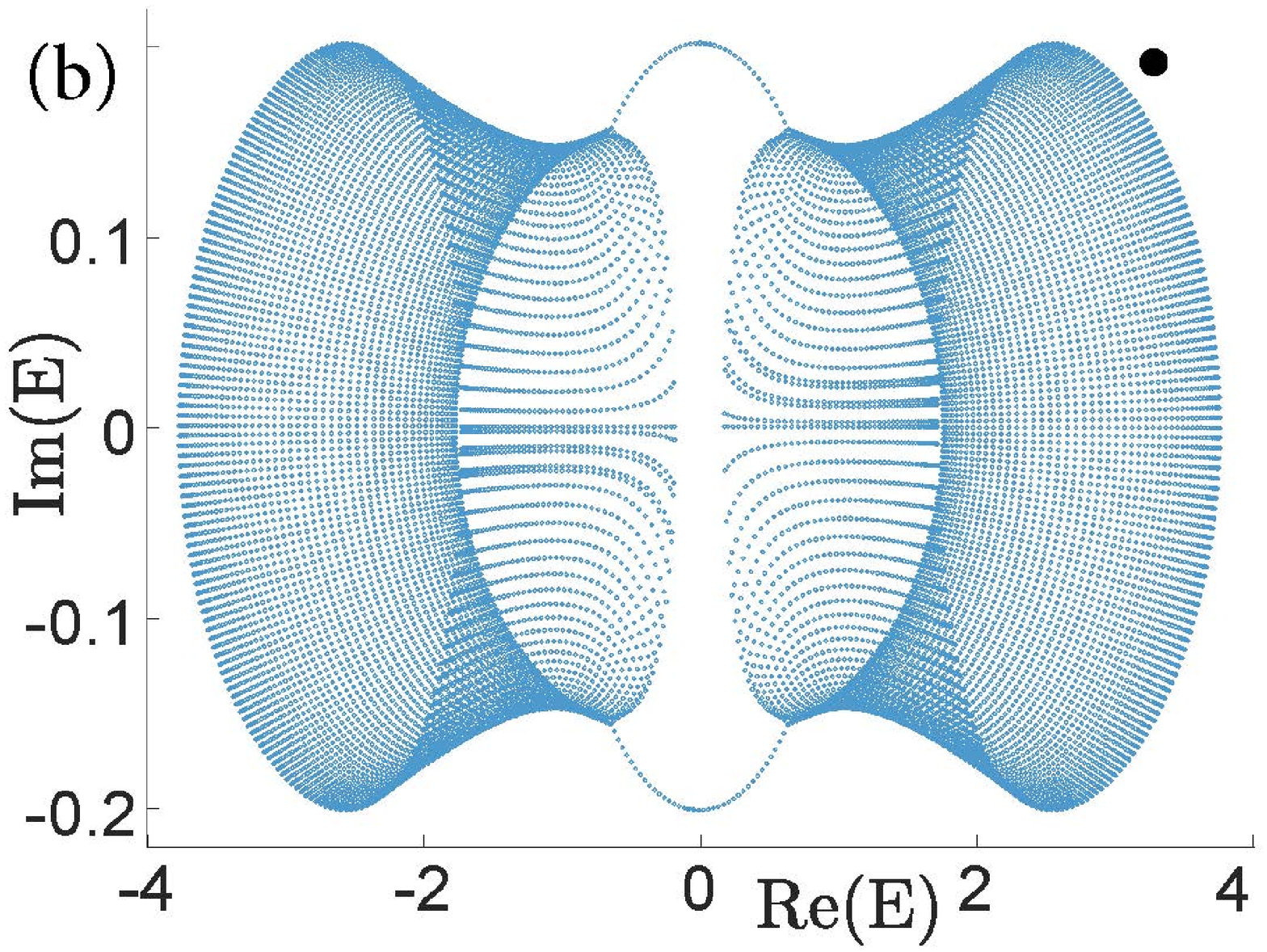}}
\caption{(a) The cylinder ``phase diagram'' of the model in Eq.(\ref{exactlybloch}). $\gamma_x=\gamma_y=\gamma$, $t=1$. The dotted curves $m_\pm$ stand for the Bloch phase boundary. (b) The cylinder spectra for $(\gamma,m)=(0.2,1.7554)$ (indicated by $\bullet$ in (a)). According to the Bloch Hamiltonian, the spectra at $(\gamma,m)=(0.2,1.7554)$ should be gapless. The spectra are in fact gapped (i.e., with two separable bands), which is consistent with the non-Bloch theory. The existence of chiral edge modes in the spectra is in accordance with $C_y=1$. }\label{exactcylinder}
\end{figure}

Finally, we discuss the cylinder topology. Suppose that the cylinder has open-boundary in the $y$ direction, hence $k_x$ remains a good quantum number. The real-space Hamiltonian can be expressed as
\begin{equation}
\begin{aligned}
\hat{H}(k_x)=&\sum_y c^\dagger_y [(m-t\cos(k_x+i\gamma_x)\sigma_z+\sin(k_x+i\gamma_x)\sigma_x] c_y \\&+\sum_y c^\dagger_y \frac{1}{2}\exp(-\gamma_y)(-i\sigma_y-t \sigma_z) c_{y+1}\\
&+\sum_y c^\dagger_y \frac{1}{2}\exp(\gamma_y)(i\sigma_y-t \sigma_z) c_{y-1}.
\end{aligned} \label{}
\end{equation}
Let us take $t=1$ for concreteness. It is convenient to solve the real-space Hamiltonian after a basis change $(\sigma_x,\sigma_y,\sigma_z)\rw (-\sigma_z,\sigma_y,\sigma_x)$. Similar to Eq.(\ref{bulkeigen2}), we obtain a quadratic equation
\begin{equation}
\begin{aligned}
&\beta^2 e^{-\gamma_y}[m-\cos(k_x+i\gamma_x)]+ e^{\gamma_y}[m-\cos(k_x+i\gamma_x)] \\&+\beta\left(E^2-\sin^2(k_x+i\gamma_x)-1-[m-\cos(k_x+i\gamma_x)]^2\right)=0,
\end{aligned} \label{}
\end{equation}
whose two solutions $\beta_{1,2}(E)$ have to satisfy $|\beta_1/\beta_2|=1$ if $E$ belongs to the continuum spectra [see discussions below Eq.(\ref{equalmod})]; therefore, Vieta's formula tells us that  \bea |\beta_{1,2}| = \exp(\gamma_y). \eea This indicates the non-Hermitian skin effect when $\gamma_y\neq 0$. Therefore, we should take $k_y\rw\tilde{k}_y-i\gamma_y$ and
the non-Bloch ``cylinder Hamiltonian'' is
\begin{equation}
\begin{aligned}
\tilde{H}_y(k_x,\tilde{k}_y)&=H(k_x\to k_x,k_y \to \tilde{k}_y-i\gamma_y)\\&=\sin(k_x+i\gamma_x)\sigma_x+\sin\tilde{k}_y \sigma_y\\&+[m- \cos(k_x+i\gamma_x)- \cos \tilde{k}_y ]\sigma_z.
\end{aligned} \label{exactHy}
\end{equation}
Its energies are
\bea && E_\pm(k_x,\tilde{k}_y) \nn\\ &&= \pm\sqrt{\sin^2(k_x+i\gamma_x)  + \sin^2\tilde{k}_y +[m- \cos(k_x+i\gamma_x)- \cos \tilde{k}_y ]^2 }.  \quad\quad \eea
As $m$ is tuned, gapless points can be created or annihilated, causing transitions between the gapped and gapless phases. Suppose that these creations/annhilations at $(k_x,\tilde{k}_y)=(0,0)$ (numerically confirmed), we have \bea m= 1+e^{\gamma_x},\, 1+e^{-\gamma_x}. \eea which are the two phase boundaries shown as solid curves in Fig.\ref{exactcylinder}(a). They differ from the Bloch phase boundary $m=m_\pm$ shown as dotted lines.

The non-Bloch cylinder Chern number is defined in the $(k_x,\tilde{k}_y)$ parameter space:  \bea C_{y(\alpha)}=\frac{1}{2\pi i}\int dk_x d\tilde{k}_y\,\epsilon^{ij} \bra{\partial_i u_{\text{L}\alpha}(k_x,\tilde{k}_y)} \partial_j u_{\text{R}\alpha}(k_x,\tilde{k}_y) \ra, \eea where $\alpha$ is the band index. We focus on the $\text{Re}(E_\alpha)<0$ band and omit the $\alpha$ index. We have numerically calculated it from Eq.(\ref{exactHy}), which gives $C_y=1$ and $C_y=0$ in the two gapped regions, respectively (indicated in Fig.\ref{exactcylinder}(a)).

\end{document}